\documentclass[sn-mathphys,Numbered]{sn-jnl}

\usepackage{graphicx}
\usepackage{float}
\usepackage{multirow}%
\usepackage{amsmath,amssymb,amsfonts}
\usepackage{amsthm}
\usepackage{mathrsfs}
\usepackage[title]{appendix}
\usepackage{xcolor}
\usepackage{textcomp}
\usepackage{manyfoot}%
\usepackage{booktabs}%
\usepackage{algorithm}
\usepackage{mathtools}
\usepackage{algorithmicx}%
\usepackage{algpseudocode}%
\usepackage{listings}%
\usepackage{lmodern}
\usepackage[capitalize]{cleveref}
\usepackage{anyfontsize}
\usepackage{hyperref}
\usepackage{cleveref}

\begin{document}

\title{Novel exact black hole solution in Dehnen $\left(1,4,\frac32\right)$ halo: thermodynamics, photon circular motion and eikonal quasinormal modes}

\author[1]{David Senjaya} \email{davidsenjaya@protonmail.com} 
\affil[1]{Department of Physics, Faculty of Science, Mahidol University, Bangkok 10400, Thailand}

\author[2]{Thanaporn Chuensuksan}
\email{thanapornchnsksn@gmail.com}
\affil[2]{Department of Physics, Faculty of Science, Srinakharinwirot University, Bangkok 10110, Thailand}

\author[3]{Supakchai Ponglertsakul}
\email{supakchai.p@gmail.com}
\affil[3]{Strong Gravity Group, Department of Physics, Faculty of Science, Silpakorn University, Nakhon Pathom 73000, Thailand}


\maketitle

\begin{abstract}

Dehnen $(1,4,\frac{3}{2})$ dark matter halo has been proven to be a valuable model for describing the surface brightness distributions of elliptical galaxies, yet its implications for black hole spacetimes remain largely unexplored. In this work, we construct a novel exact black hole solution embedded in this Dehnen halo and investigate its physical consequences. The influence of the halo on black hole thermodynamics is analyzed through the mass function, entropy, Hawking temperature, heat capacity, and Gibbs free energy, allowing us to assess both local and global thermodynamic stability of the black hole–dark matter system. Our results show that the presence of a Dehnen-type halo not only stabilizes the otherwise thermodynamically unstable Schwarzschild black hole but also induces phase transitions. In addition, we study null geodesics to examine photon motion, the shadow radius and the optical appearance of the system. The dark matter halo modifies the effective potential, leading to observable changes in the photon sphere and the apparent size of the shadow. We also explore the instability of circular null geodesics and its relation to quasinormal modes in an eikonal limit. These findings highlight the significant role of realistic dark matter distributions in shaping both the thermodynamic behavior and the observable signatures of black holes, providing further insight into the interplay between dark matter halos and central black holes in galaxies.

\end{abstract}

\section{Introduction}\label{sec: intro}
Despite extensive testing, Einstein's General Relativity cannot fully explain black hole singularities, dark matter and dark energy. Although gravitational waves and black holes' shadow images from the Event Horizon Telescope \cite{LIGOScientific:2016vlm,EventHorizonTelescope:2019ggy,EventHorizonTelescope:2022wkp} confirm the prediction of black holes, the theory is still considered incomplete. Among the remaining problems in general relativity, dark matter stands as one of the most profound and persistent mysteries, challenging our understanding of the universe. Its existence can be inferred from various astrophysical and cosmological observations, including galaxy rotation curves \cite{Rubin:1970zza}, gravitational lensing \cite{Massey:2010hh}, the cosmic microwave background \cite{Planck:2018vyg} and collision of galaxy clusters \cite{Clowe:2003tk,Markevitch:2003at}.

On the galactic scales, dark matter plays a crucial role in stellar dynamics around galaxies. For instance, the fact that galactic rotation curves stay flat for big galaxies prompted the dark matter theory to explain the invisible mass. An astronomical research reveals that the rotating movements of stars in these galaxies can only be explained by dark matter \cite{Gebhardt:2011yw}, which might account for up to 90 percent \cite{Mathew2014Essays,fidler2018dark}  of a galaxy's total mass, whereas the remainder made up of conventional baryonic matter. Moreover, during the early phases of the universe, dark matter collected around galactic centers to promote star formation. As the galaxies developed to reach their mature stage, dark matter drifted outward, generating halos. An advancement of astrophysical instruments reveals that there are evidences that most galaxies are surrounded by dark matter halo stretching far beyond a visible part of galaxies \cite{Wechsler:2018pic}. Moreover, recent observations have confirmed existence of supermassive black hole at the centre of galaxy $M87^\ast$ \cite{EventHorizonTelescope:2019ggy} and the Milky Way galaxy \cite{EventHorizonTelescope:2022wkp}.

Black holes embedded in dark matter environments provide a natural laboratory to explore the mutual interaction between black holes and dark matter. With the presence of dark matter, the gravitational field of black hole is modified. This could potentially leave imprints on observable quantities such as black hole shadows, gravitational lensing and the motion of test particles. In recent years, there has been many interests in finding black hole solutions embedded in various dark matter halo profiles. A rotating black hole surrounded by Universal Rotation Curve dark matter profile is constructed. Shadow's radius of such a Kerr-like solution is found to be smaller than those of the Kerr black hole \cite{PhysRevD.100.044012} given the core radius is comparable to the mass of black hole and very high core's density. By using several dark matter profiles such as Navarro-Frenk-White, Burkurt and Hernquist as source of energy-momentum tensor, the authors of \cite{Konoplya:2022hbl} solve the Einstein field equation and find the spacetime metrics representing black holes embedded in galactic halos. For cold dark matter Einasto profile, exact and numerical black hole solutions are explored in \cite{Figueiredo:2023gas,Liu:2023oab}. Shadow and gravitational lensing of Schwarzschild black hole embedded in the Hernquist profile is significantly different from the Schwarzschild vacuum solution \cite{Xavier:2023exm,Jha:2025xjf}. More interestingly, a suitable incorporation of dark matter halos allows one to construct  regular black hole solutions \cite{Kar:2025phe}.

A Dehnen profile is flexible model which can give rise to either cored or cuspy profile. It was originally developed to describe mass distribution in galaxies \cite{dehnen}. In particular, it is commonly applied to dwarf galaxies. In general, dwarf galaxies do not host black hole at their centre. But some recent observations indicate a possibility of supermassive black holes hosted by $Mrk~462, Henize~2-10$ and $Leo~I$ \cite{mrk462,Henize,Leo}. A particular class of dwarf galaxy called ultra fainted dwarf galaxies usually contain very old and long-lived stars. Therefore, it is one of the least luminosity known galaxy. Consequently, it has the highest dark matter to light ratio. Hence, it is a perfect place to explore nature of dark matter. In 2022, rotating black hole immersed in the Dehnen halo is constructed and its shadow properties are investigated \cite{Pantig:2022whj}. To distinguish dark matter profile whether it is cored or cuspy, weak deflection angle provides a much better way to differentiate dark matter profile as reported in \cite{Pantig:2022whj}. This work has sparked interest and led to many studies devoted to black hole solutions surrounded by various Dehnen dark matter profiles \cite{Gohain:2024eer,Al-Badawi:2024asn,Toshmatov:2025rln,Uktamov:2025lwb,Senjaya:2025via}. 

In this work, we use the approach introduced in \cite{Xu:2018wow} to derive a new static spherically symmetric black hole solution immersed in the Dehnen-type $\left(1,4,\frac{3}{2}\right)$ density profile describing dark matter halo. The Dehnen $\left(1,4,\tfrac{3}{2}\right)$ profile has proven to be a valuable tool in modeling the surface brightness distribution of elliptical galaxies \cite{Al-Badawi:2024asn,Rani:2025esb}. Specifically, the choice of $\gamma=3/2$ (see \eqref{dehnenprofile} below) shows the most resemblance to the de Vaucouleurs's surface brightness profile \cite{dehnen}. Despite its success in capturing the observed stellar density structures, a corresponding analytical black hole solution within this framework remains unexplored. Investigating such a solution is essential, as it would provide deeper insights into the interplay between galactic density profiles and central compact objects, thereby bridging a critical gap between astrophysical modeling and gravitational theory.

This article is organized as follow. In section \ref{sec:BH+DM}, we construct an exact static spherically symmetric black hole surrounded by the Dehnen $\left(1,4,\frac{3}{2}\right)$ density profile. Then, the singularity structure and energy conditions are investigated. Thermodynamics quantities such as temperature, heat capacity and free energy of the solution in section \ref{sec:thermo}. In section \ref{sec:null}, we analyze geodesics of massless particle around the black holes. Black hole's shadow radius is computed and compared with observational data. In section \ref{sec:Lyapunov}, we explore instability of circular null orbit via Lyapunov exponent. Quasinormal frequencies in eikonal limit are derived in section \ref{sec:QNMs}. We summarize our findings in the last section \ref{sec: conclu}.

\section{Black hole + DM profile}\label{sec:BH+DM}
To construct the spacetime metric describing black hole immersed in dark matter halo, we follow the method that is outlined in \cite{Xu:2018wow}. We start by considering a static spherically symmetric black hole surrounded by dark matter halo. The general metric is given by 
\begin{align}
    ds^2 &= -\left[f(r)+F(r)\right]dt^2 + \left[g(r) + G(r)\right]^{-1}dr^2 + r^2\left(d\theta^2+\sin^2\theta d\phi^2\right), \label{1stmetric}
\end{align}
where throughout this work, we have set $c=G=1$. In an absence of black hole, i.e., $F=G=0$, the spacetime background above reduces to pure dark matter spacetime. The energy-momentum tensor solely comes from the presence of dark matter halo. Therefore, the Einstein field equation can be written as
\begin{align}
   {R_a}^b+\frac{1}{2}R{\delta_a}^b &= 8\pi {T_a}^b,\label{EFE}
\end{align}
where ${T_a}^b=\operatorname{diag}\left[-\rho_E,p_r,p,p\right].$ represents energy momentum tensor of dark matter. The independent components of the Einstein field equation are
\begin{align}
    8\pi{T_t}^t &= \frac{\left(-1+g+G\right)}{r^2} + \frac{\left(g+G\right)'}{r}, \label{temporaleq}\\
  8\pi  {T_r}^r &= \left(g+G\right)\left[\frac{\left(f+F\right)'}{r\left(f+F\right)} + \frac{1}{r^2}\right] - \frac{1}{r^2}, \label{radialeq}\\
  8\pi  {T_\theta}^\theta &= \frac{\left(g+G\right)}{2\left(f+F\right)}\left[\left(f+F\right)'' -\frac{{\left (f+F\right)'}^2}{2\left(f+F\right)} + \frac{\left(f+F\right)'}{r}\right] \nonumber \\
    &~~~~+ \left(g+G\right)'\left[\frac{\left(f+F\right)'}{4\left(f+F\right)}+\frac{1}{2r}\right],\label{angulareq}
\end{align}
where prime denotes derivative with respect to $r$. Without dark matter halo, i.e., vanishing energy momentum tensor and $f=g=1$, the above equations admit the Schwarzschild solution. Thus, we obtain
\begin{align}
    F &= G= -\frac{r_s}{r},
\end{align}
where $r_s = 2M$ with $M$ being the Schwarzschild mass. 

In this work, we are interested in the Dehnen type dark matter halo \cite{dehnen}, having the following form of mass density, 
\begin{equation}
    \rho(r)=\rho_0\left(\frac{r}{r_0}\right)^{-\gamma}\left(1+\left(\frac{r}{r_0}\right)^\alpha\right)^{\frac{\gamma-\beta}{\alpha}}. \label{dehnenprofile}
\end{equation}
We specifically choose $(\alpha,\beta,\gamma)=\left(1,4,\frac{3}{2}\right)$ such that the Dehenen profile matches the de Vaucouleurs's law ($r^{1/4}$ law). This yields the following explicit expression, 
\begin{equation}
    \rho(r)=\rho_0\left(\frac{r}{r_0}\right)^{-\frac{3}{2}}\left(1+\frac{r}{r_0}\right)^{-\frac{5}{2}} \label{rho},
\end{equation}
where $\rho_0$ is a constant acting as the dark matter halo central density parameter and $r_0$ is halo core radius \cite{Gohain:2024eer}. In spherically symmetric spacetime, the mass distribution function can be calculated by \cite{MovanBook} 
\begin{align}
    M_{DM} &= 4\pi\int_0^r \rho(\hat{r})\hat{r}^2 d\hat{r} = \frac{8}{3}\pi\rho_0r_0^3\left(1+\frac{r_0}{r}\right)^{-\frac{3}{2}}.   \label{massdist} 
\end{align}

It is clear that the dark matter mass profile depends on the central density $\rho_0$ and halo's core radius $r_0$. A test particle moving under an influence of dark matter has a tangential velocity that relates to the mass distribution \eqref{massdist}
\begin{align}
    v_t(r)&=\sqrt{\frac{M_{DM}(r)}{r}}=\sqrt{\frac{8}{3}\pi\rho_0r_0^3}\left(1+\frac{r_0}{r}\right)^{-\frac{3}{4}}r^{-\frac{1}{2}}.
\end{align}
In a pure dark matter spacetime, i.e., $F=G=0$, the tangential velocity relates to the red shift function of the spacetime metric $f$ \cite{Xu:2018wow}
\begin{align}
v_t^2(r) &= r \frac{d}{dr}\left( \ln\sqrt{f(r)}\right),
\end{align}
where we impose $f=g$. Thus, we can solve for the redshift function
\begin{align}
 f(r)&=e^{2\int \frac{v_t^2(r)}{r}dr },\nonumber\\
 &=e^{\frac{32}{3}\pi\rho_0r_0^2\left(1+\frac{r_0}{r}\right)^ {-\frac{1}{2}}}. 
\end{align}
Note that, the central density $\rho_0$ can be re-written in term of dark matter mass $M_{DM}$ via \eqref{massdist}. 
Therefore, a static spherically symmetric spacetime describing a black hole surrounded by a dark matter halo with Dehnen type $(1,4,\frac{3}{2})$ can be written explicitly as
\begin{align}
    ds^2 &= -\left( e^{4M_{DM}\left(\frac{1}{r}+\frac{1}{r_0}\right)} - \frac{r_s}{r}\right)dt^2 + \frac{dr^2}{\left( e^{4M_{DM}\left(\frac{1}{r}+\frac{1}{r_0}\right)} - \frac{r_s}{r}\right)} + r^2\left(d\theta^2+\sin^2\theta d\phi^2\right). \label{metric}
\end{align}
We notice that $g_{tt} \to e^{4M_{DM}/r_0}$ as $r\to \infty$, which is clearly constant. Moreover, in the absence of dark matter $M_{DM}=0$, the metric above reduces to the standard form of the Schwarzschild metric. For theoretical purposes, we shall fix black hole mass $M=1$ throughout this study, unless otherwise stated.
\begin{figure}[h]
    \centering
    \includegraphics[scale=0.35]{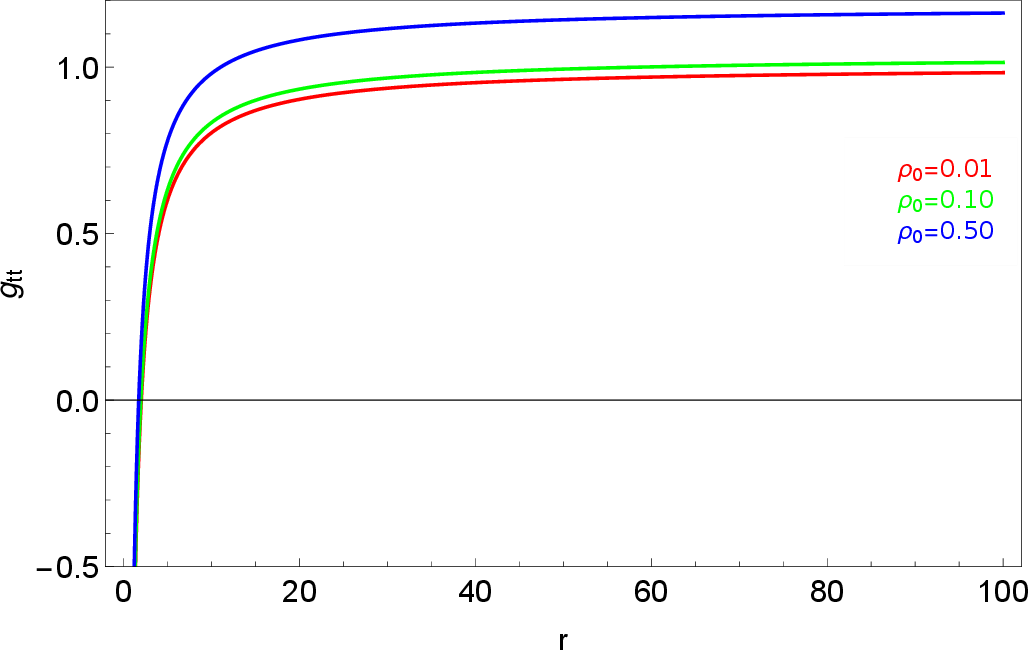}
    \includegraphics[scale=0.35]{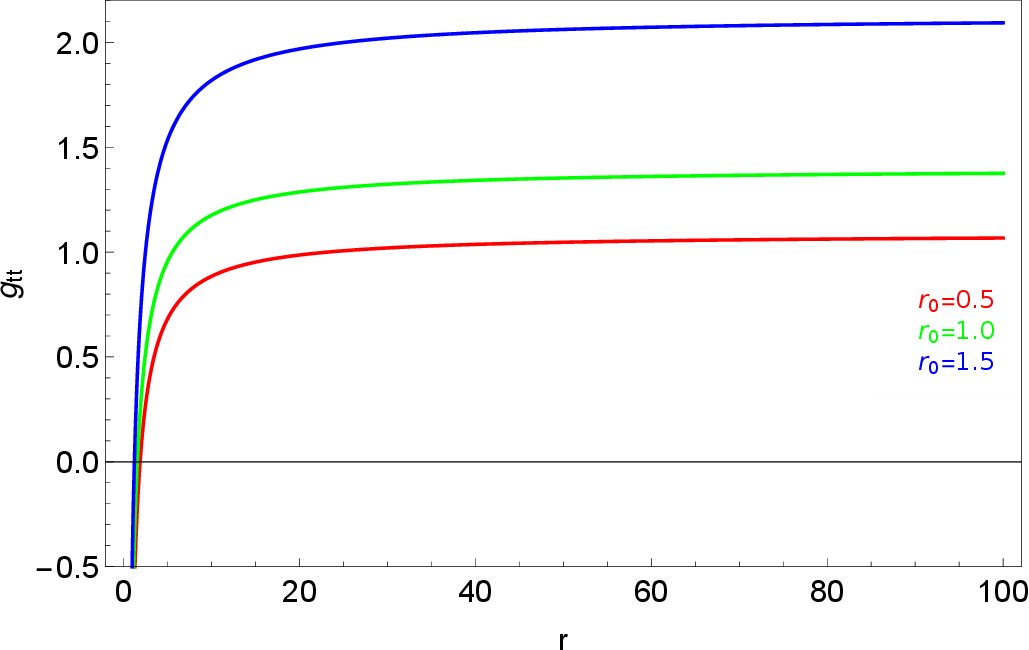}
    \caption{Metric function $g_{tt}$ as a function of $r$ for $r_s=2$. Left: fixed $r_0=0.1$, Right: fixed $\rho_0=0.01.$} \label{fig:metric}
\end{figure}

\begin{figure}[h]
    \centering
    \includegraphics[scale=0.32]{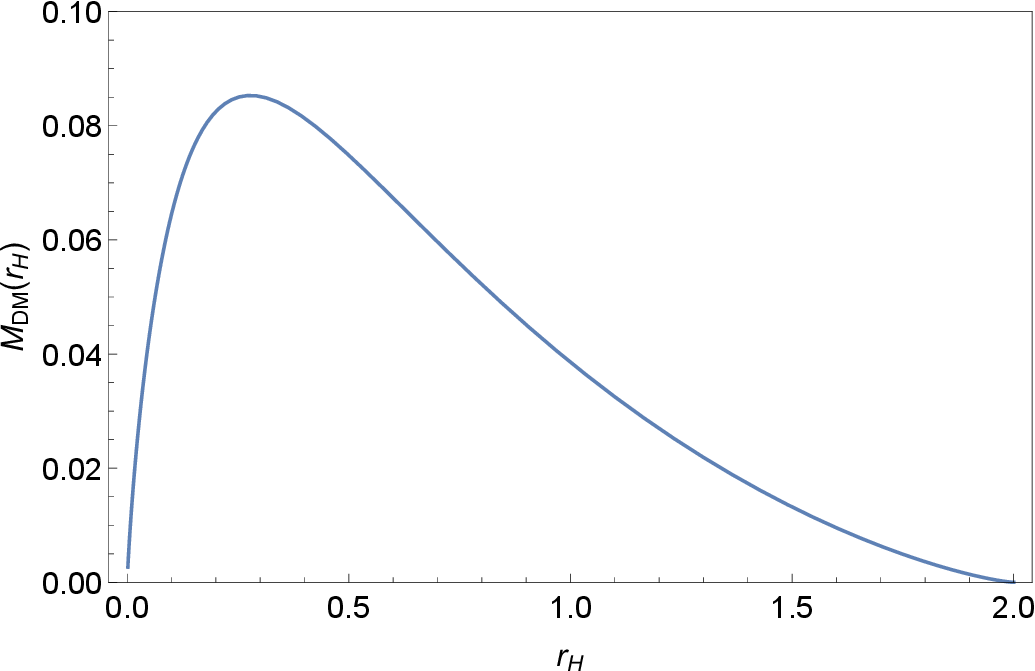}
    \includegraphics[scale=0.5]{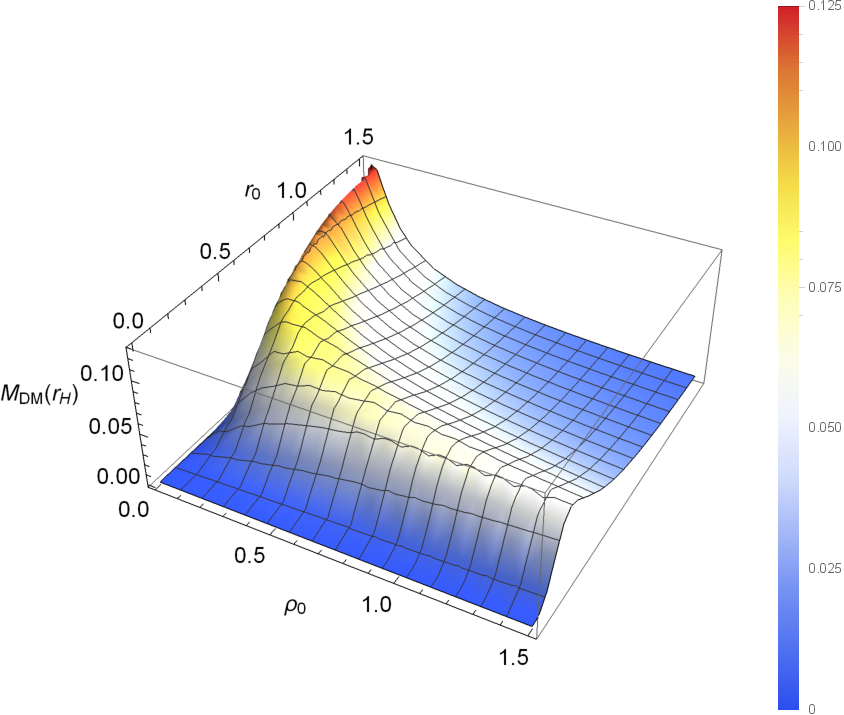}
    \caption{Left: $M_{DM}(r_H)$ as a function of $r_H$. Right: 3D plot of $M_{DM}(r_H)$ as a function of $\rho_0$ and $r_0$. } \label{fig:MDM}
\end{figure}

In Fig.~\ref{fig:metric}, we display behavior of the metric function $g_{tt}$ as a function of $r$. It can be clearly seen that $g_{tt}$ approaches a constant value at large $r$. The location of the black hole's event horizon can be determined from the zero of $g_{tt}$. For instance, the event horizon locates at $r=r_H=1.857$ for $\rho_0=0.01$ and $r_0=0.5$. 

The left panel of Fig.~\ref{fig:MDM} demonstrates the value of the dark matter halo mass at the horizon $M_{DM}(r_H)$ plotting against the radius of the horizon $r_H$. Recall that, $M_{DM}(r_H)$ and $r_H$ depend explicitly on $\rho_0,r_0$. Therefore, for the sake of demonstration, we set $\rho_0=r_0$ in this plot. We observe that the dark matter mass reaches its maximum at $r_H=0.27$ and $M_{DM}(r_H)=0.085$. After $r_H>0.27$, the dark matter halo mass at the horizon decreases monotonically with $r_H$. In the right panel of Fig.~\ref{fig:MDM}, we depict how $M_{DM}(r_H)$ is influenced by the central dark matter energy density $\rho_0$ and the core's radius $r_0$. Interestingly, the maximum mass attains at small $\rho_0$ and large $r_0$.

\begin{figure}[h]
    \centering
    \includegraphics[scale=0.24]{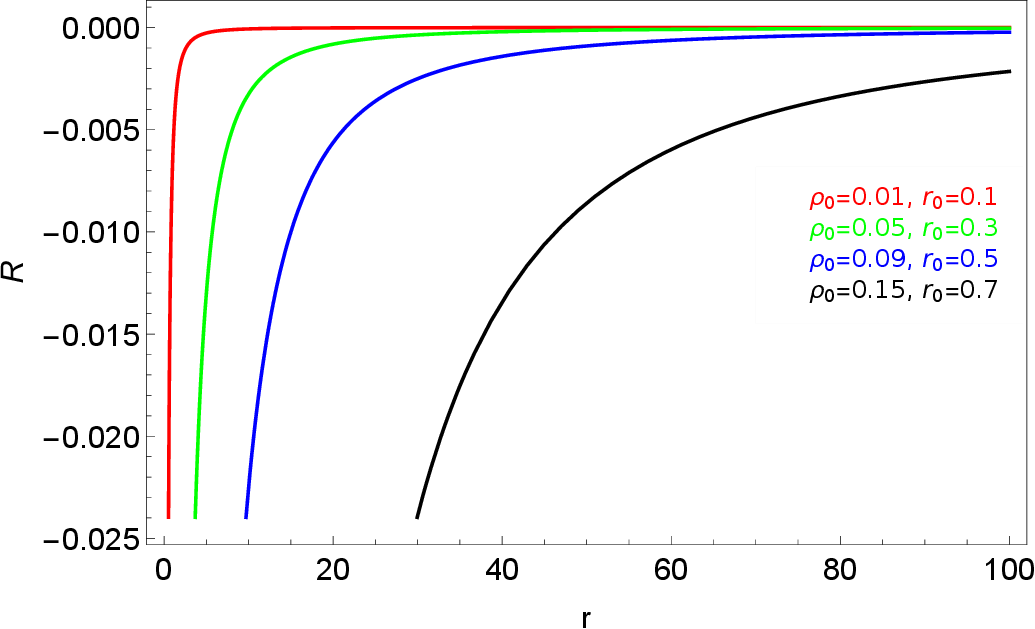}
    \includegraphics[scale=0.24]{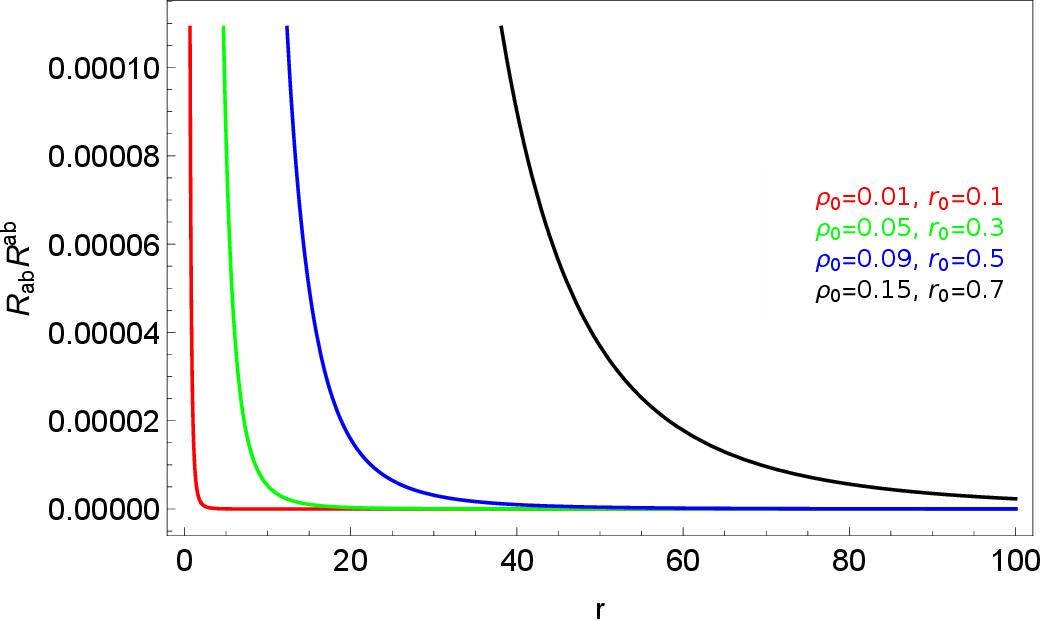}   
    \includegraphics[scale=0.24]{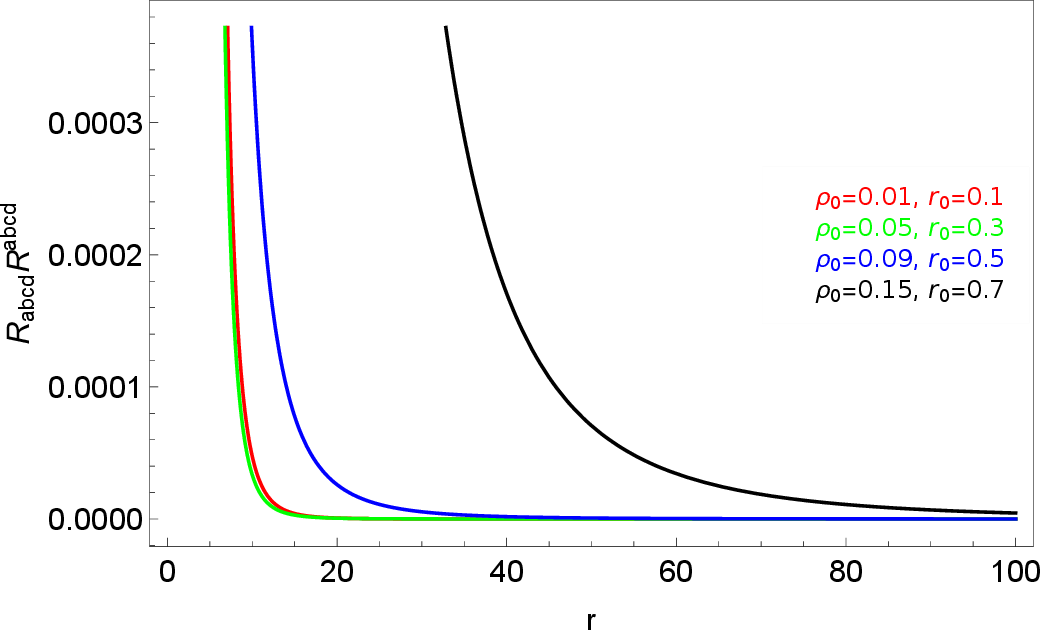}
    \caption{Graphical demonstration of three curvature invariants Left: $R$, Middle: $R_{ab}R^{ab}$ and Right: $R_{abcd}R^{abcd}$.} \label{fig:invcalar}
\end{figure}

Now let us investigate the existence of singularity with the presence of dark matter halo. Here, we explore the Ricci scalar $(R)$, Ricci scalar square $R_{ab}R^{ab}$ and the Kretschmann scalar $R_{abcd}R^{abcd}$. For the given metric \eqref{1stmetric}, these are explicitly obtained
\begin{align}
    R &= \frac{2\left(1-f-F\right)}{r^2} - \frac{4\left(f+F\right)'}{r} - \left(f+F\right)'', \\
    R_{ab}R^{ab} &= 4\left[ \frac{\left(f+F-1\right)^2}{2r^4}+ \frac{\left(f+F-1\right)\left(f+F\right)'}{r^3} + \frac{{\left(f+F\right)'}^2}{r^2} \right. \nonumber \\
&\left.~~~~~~~~~+ \frac{\left(f+F\right)'\left(f+F\right)''}{2r} + \frac{{\left(f+F\right)''}^2}{8}\right],\\   
    R_{abcd}R^{abcd} &= \frac{4\left(f+F-1\right)^2} {r^4} + \frac{4{\left(f+F\right)'}^2} {r^2} + {\left(f+F\right)''}^2.
\end{align}
These expressions resemble those of the Schwarzschild case when $f=g=1$ i.e., $\{R,R_{ab}R^{ab},R_{abcd}R^{abcd}\} = \{0,0,\frac{12rs^2}{r^6}\} $. Moreover, we illustrate three curvature invariants as radial function in Fig.~\ref{fig:invcalar}. It is clear that all curvature scalars diverge as $r\to 0$ and are finite for $r>0$. This indicates that the spacetime metric \eqref{metric} possesses an essential singularity, while the event horizon $r_H$ is merely a coordinate singularity. In addition, we observe that the Ricci scalar is always negative which infers that the spacetime considered here is non-Ricci flat.

To gain a deeper understanding of the spacetime background \eqref{1stmetric}, we shall consider energy conditions. First, we obtain the energy momentum tensor ${T_\mu}^\nu=\operatorname{diag}\left[-\rho_E,p_r,p,p\right]$ from the Einstein field equations \eqref{temporaleq}-\eqref{angulareq}
\begin{align}
    \rho_E &= -p_r = \frac{1}{8\pi r^2} - \frac{e^{4M_{DM}\left(\frac{1}{r}+\frac{1}{r_0}\right)}\left(r+2M_{DM}\right)}{8\pi r^3}, \label{rhoE} \\
    p &= \frac{M_{DM}e^{4M_{DM}\left(\frac{1}{r}+\frac{1}{r_0}\right)}}{16\pi r^4}\left[\frac{3rr_0+4M_{DM}\left(r+r_0\right)}{\left(r+r_0\right)}\right]. \label{p}
\end{align}
With the presence of dark matter halo, we probe four energy conditions as a result of spacetime metric \eqref{metric} i.e., null, weak, dominant, and strong energy conditions. 

The null energy condition (NEC) states that for all null vectors $n^a$, the energy momentum tensor must satisfy 
\begin{align}
    T_{ab}n^a n^b &\geq 0.
\end{align}
This puts the following constraint on the energy density and pressure \cite{Uktamov:2025lwb,Toshmatov:2017kmw}
\begin{align}
    \rho_E + p_{j} \geq 0, 
\end{align}
where $j=1,2,3$ and $p_1=p_r, p_2=p_3=p$. From \eqref{rhoE}, it is obvious that $\rho_E+p_r = 0$. Hence, the NEC solely depends on $\rho_E+p\geq 0$. This reads explicitly
\begin{align}
    \frac{1}{8\pi r^2} + e^{4M_{DM}\left(\frac{1}{r}+\frac{1}{r_0}\right) }  \left[ \frac{M_{DM}^2}{4\pi r^4} - \frac{1}{8\pi r^2} - \frac{M_{DM}\left(4r+r_0\right)}{16\pi r^3 \left(r+r_0\right)}\right]&\geq 0. \label{NEC}
\end{align}

The weak energy condition (WEC) implies that the total energy density as measured by any time-like observer $(u^a)$ is always positive. Mathematically, this can be written as
\begin{align}
    T_{ab}u^au^b &\geq 0.
\end{align}
It can be translated into the constraints on $\rho_E,p_r$ and $p$ as \cite{Uktamov:2025lwb,Toshmatov:2017kmw}
\begin{align}
    \rho_E &\geq0,~~~~~~\rho_E+p_j \geq 0. \label{WEC}
\end{align}
This adds one additional constraint to the NEC. Therefore, if the WEC is satisfied, the NEC will simultaneously hold. However, the energy density, $\rho_E$ \eqref{rhoE}, is never positive. This is because the only positive term is $\frac{1}{8\pi r^2}$ which will always be subjugated by $-\frac{e^{4M_{DM}\left(\frac{1}{r}+\frac{1}{r_0}\right)}}{8\pi r^2}$. Thus, only the second condition of the WEC might possibly be satisfied. 

The physical interpretation of the dominant energy condition (DEC) is that the local energy flux cannot propagate faster than the speed of light relative to any local future-directed observer \cite{Uktamov:2025lwb}. Equivalently, this can be translated into
\begin{align}
    \rho_E - |p| &\geq 0,
\end{align}
or
\begin{align}
    \frac{1}{8\pi r^2} - e^{4M_{DM}\left(\frac{1}{r}+\frac{1}{r_0}\right) }  \left[ \frac{M_{DM}^2}{4\pi r^4} + \frac{1}{8\pi r^2} + \frac{M_{DM}\left(4r+7r_0\right)}{16\pi r^3 \left(r+r_0\right)}\right]&\geq 0. \label{DEC}
\end{align}
It turns out that the DEC is never positive in the same manner as the $\rho_E$.

Lastly, the strong energy condition (SEC) states that \cite{Uktamov:2025lwb}
\begin{align}
    \left(T_{ab}-\frac{1}{2}Tg_{ab}\right)n^a n^b &\geq 0,
\end{align}
which yields
\begin{align}
    \rho_E+p_r+2p &\geq 0.
\end{align}
Hence, we obtain
\begin{align}
  \frac{e^{4M_{DM}\left(\frac{1}{r}+\frac{1}{r_0}\right)}M_{DM}}{2\pi r^3}\left[  \frac{3r_0}{4\left(r+r_0\right)} + \frac{M_{DM}}{r}\right] &\geq 0. \label{SEC}
\end{align}
The expression on the left hand side is clearly positive for all $r$. Therefore, the SEC is generally satisfied. 

Furthermore, asymptotic behavior of these energy conditions are as follow
\begin{align}
    \lim_{r\to 0}  \{\eqref{NEC}, \eqref{WEC}, \eqref{DEC}, \eqref{SEC} \} &= \pm \infty, \\
     \lim_{r\to \infty}  \{\eqref{NEC}, \eqref{WEC}, \eqref{DEC}, \eqref{SEC} \} &= 0.
\end{align}

\begin{figure}[h]
    \includegraphics[scale=0.35]{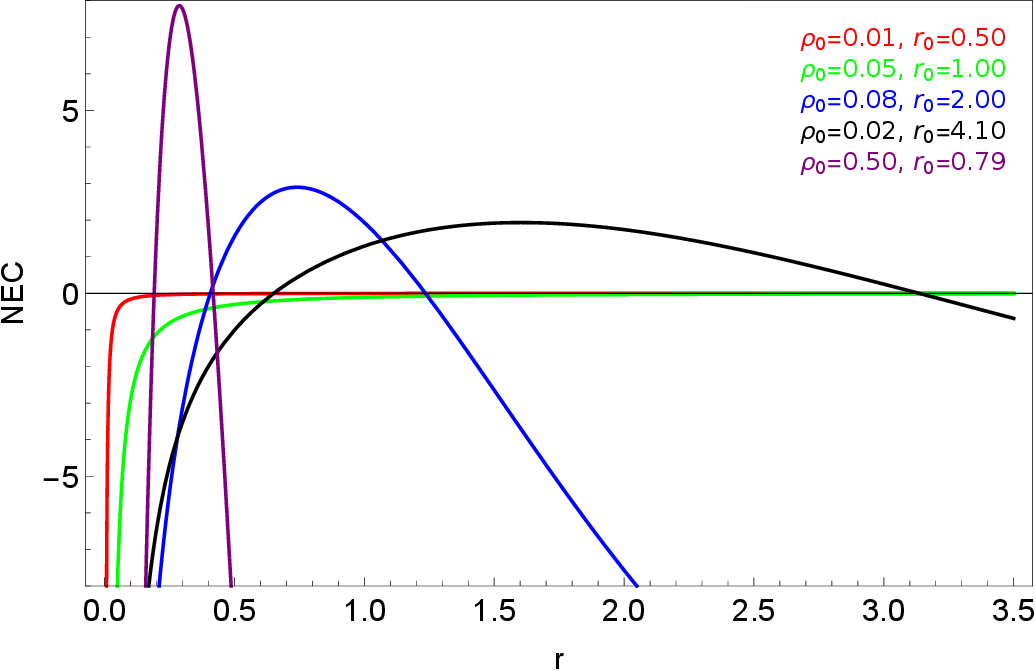}
    \includegraphics[scale=0.35]{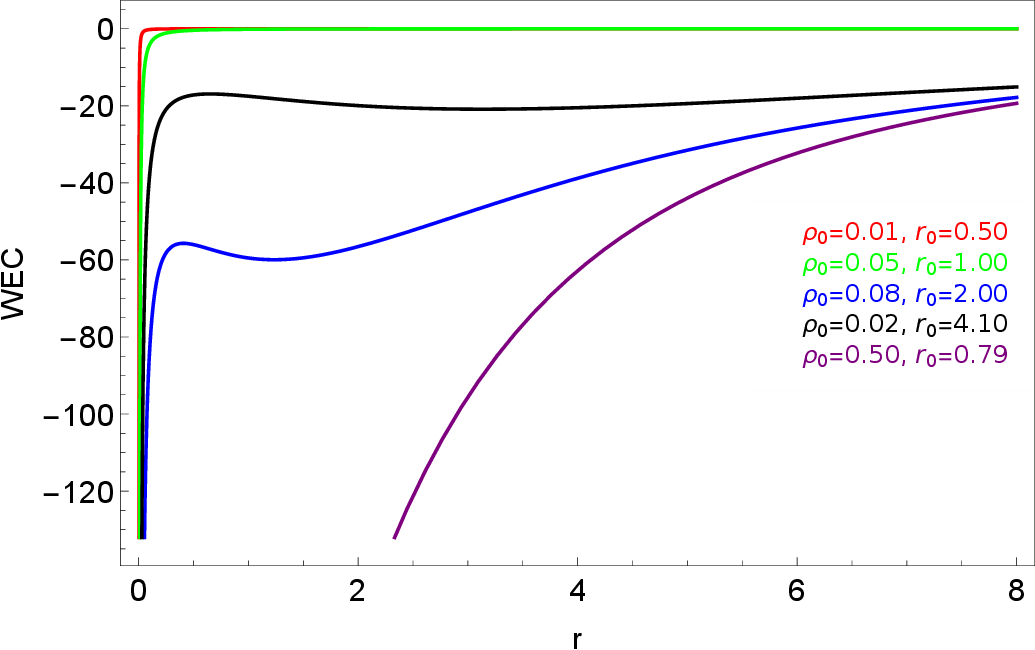}\\
    \includegraphics[scale=0.35]{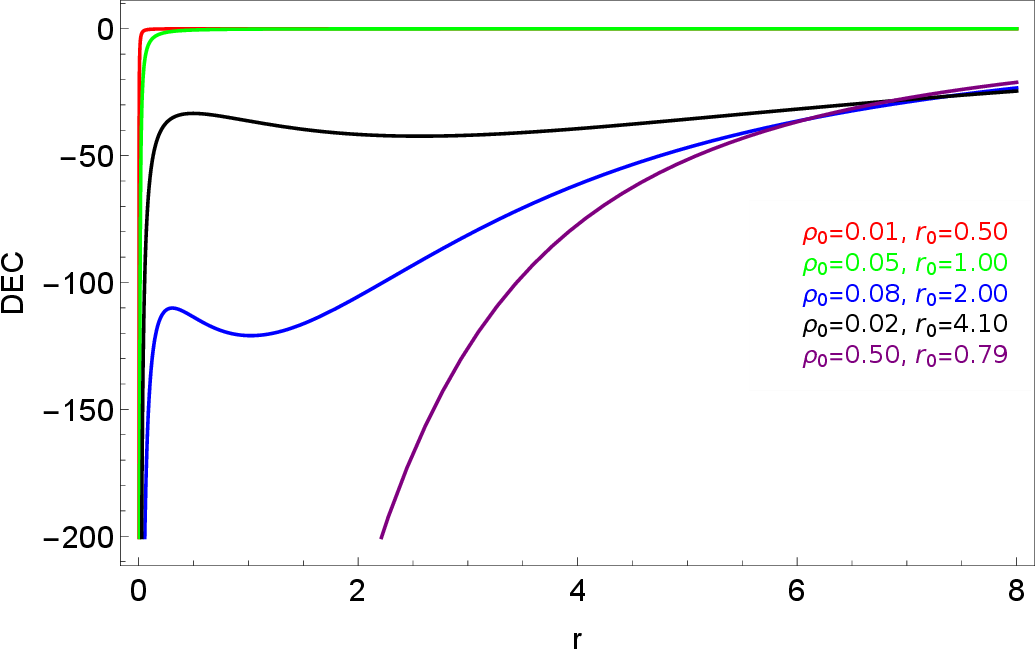}
    \includegraphics[scale=0.35]{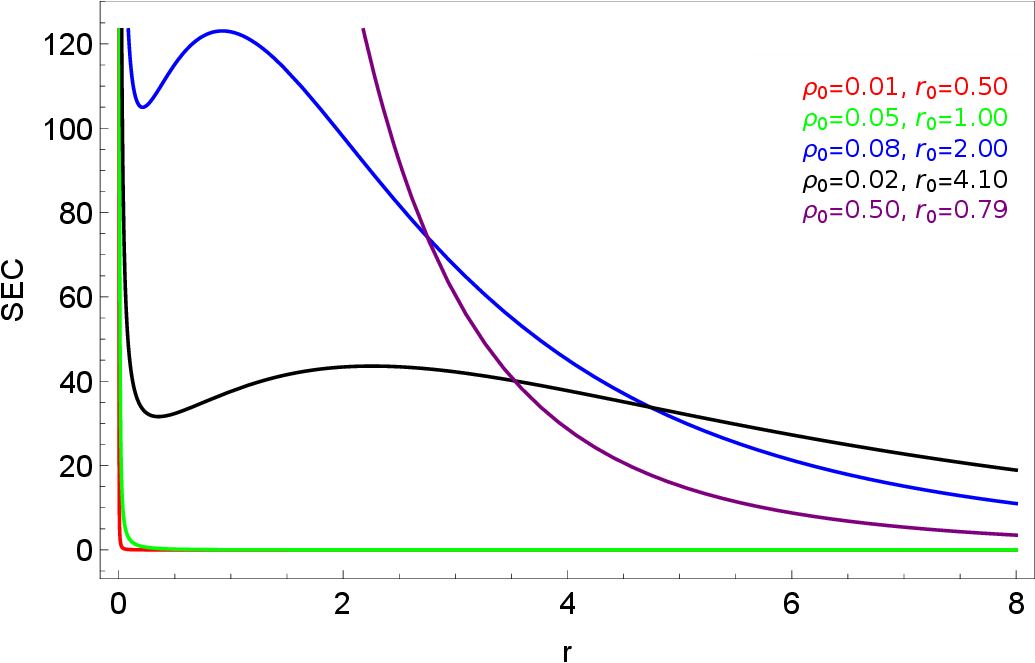}
    \caption{Four energy conditions as radial function for various value of $\rho_0$ and $r_0$. In the WEC panel, only $\rho_E\geq0$ is shown.} \label{fig:energycond}
\end{figure}

To complete our analysis, we present how each energy condition (NEC, WEC, DEC and SEC) behave as a function of the radial coordinate for fixed value of $\rho_0$ and $r_0$ in Fig.~\ref{fig:energycond}. We provide a graphical analysis of the inequalities given by \eqref{NEC}, \eqref{WEC} (only the first condition), \eqref{DEC} and \eqref{SEC}. The positive regions in these figures illustrate the regions where the respective energy conditions are satisfied. It is observed that the NEC can be satisfied if the product between $\rho_0$ and $r_0$ is sufficiently large. In addition, we find that the positive regions in the NEC panel occur outside the event horizon. However, after reaching certain radii, the NEC is no longer satisfied. The WEC and DEC are, however, violated as discussed earlier. We remark that the second condition of the WEC is satisfied similarly to the NEC panel. Finally, we find that the SEC is satisfied for all $r$. 

In the presented solution, the introduction of dark matter significantly modifies the stress–energy tensor, leading to an effective anisotropic fluid. In particular, the radial pressure $p_r$ can become positive over a substantial region of spacetime. When combined with a sufficiently small energy density or strong pressure anisotropy, this behavior results in violations of the weak and dominant energy conditions. Physically, these violations originate from the smearing of the dark matter distribution over a characteristic length scale, a mechanism that is known to regularize the geometry at the expense of classical energy conditions. Such features are common in quantum-inspired and dark-matter–inspired black hole models, i.e., while they allow for singularity-free configurations, they generically violate the WEC and/or DEC. Indeed, under physically reasonable assumptions on the mass function, all known regular and rotating black hole solutions violate the weak energy condition somewhere in the spacetime \cite{Nicolini:2005vd, Ansoldi:2006vg, Neves:2014aba, Zhang:2020mxi,Tangphati:2023xnw, Sajadi:2025prp,Khan:2025sts}. Moreover, recent studies have also reported violations of the weak and dominant energy conditions for static black holes immersed in Hernquist dark matter halos \cite{Stelea2023, Myung2025}.

\section{Black Hole Thermodynamics}\label{sec:thermo}
In this section, we investigate the thermodynamics of a Schwarzschild black hole immersed in the Dehnen $\left(1,4,\frac{3}{2}\right)$ dark matter profile. We begin by considering the position of the horizon $r_H$, which is obtained from,
\begin{equation}
    g_{tt}(r_H)=0\to e^{\frac{32}{3}\pi \rho_0 r_0^2\left(1 + \frac{r_0}{r_H}\right)^{-1/2} } - \frac{2M}{r_H}=0,
\end{equation}
or equivalently,
\begin{align}
    M &= \frac{r_H}{2} e^{\frac{32}{3}\pi \rho_0 r_0^2\left(1 + \frac{r_0}{r_H}\right)^{-1/2} } \label{mass}.
\end{align}
As one can see, in the absence of dark matter $\rho_0=0$. The above relation reduces to the Schwarzschild radius ($r_H=r_s=2M$). In black hole thermodynamics, the equivalent enthalpy of classical thermodynamics is exactly the mass $M$ as given by \eqref{mass} \cite{Kastor:2009wy,Kastor:2018cqc}. 

\begin{figure}[H]
    \centering
    \includegraphics[scale=0.4]{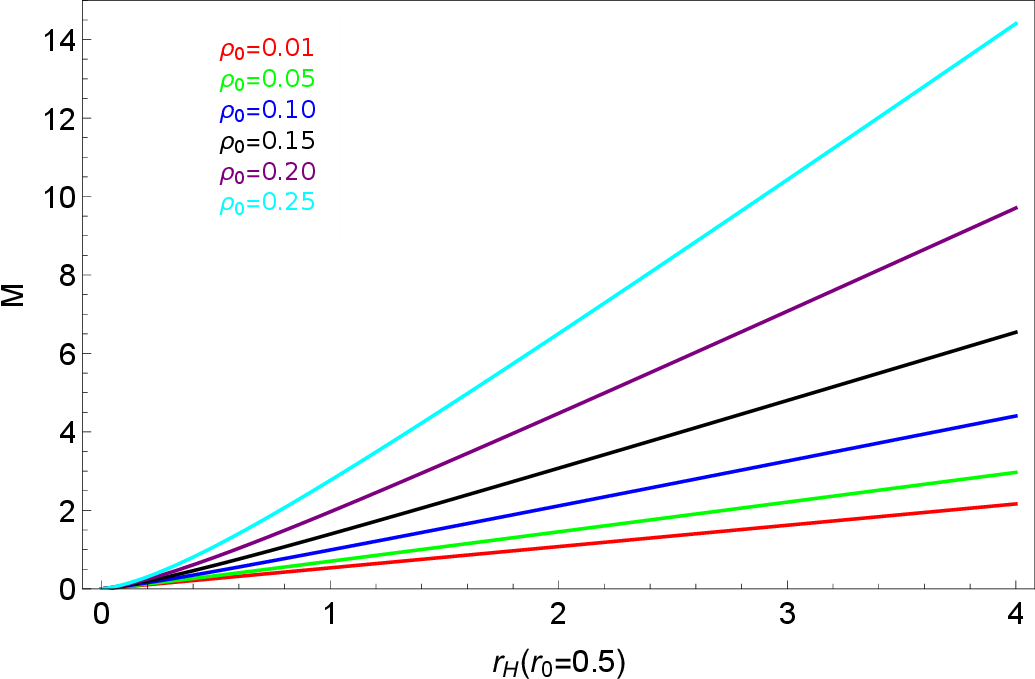}
    \caption{The mass function plotted against $r_H$ for different core density $\rho_0$ and fixed $r_0=0.5$. } 
  \label{fig:M}
\end{figure}
We illustrate the mass of black hole as a function of the event horizon $r_H$ for various central density $\rho_0$ and fixed core radius $r_0=0.5$ in Fig.~\ref{fig:M}. As expected, the mass increases with $r_H$. The more central density, the higher the rate of increasing. The mass function differs significantly as $r_H$ gets larger. A similar trend is also observed for the other value of $r_0$. However, in smaller $r_0$, the difference of each curve becomes less significant. 

\begin{figure}[H]
    \centering
    \includegraphics[scale=0.35]{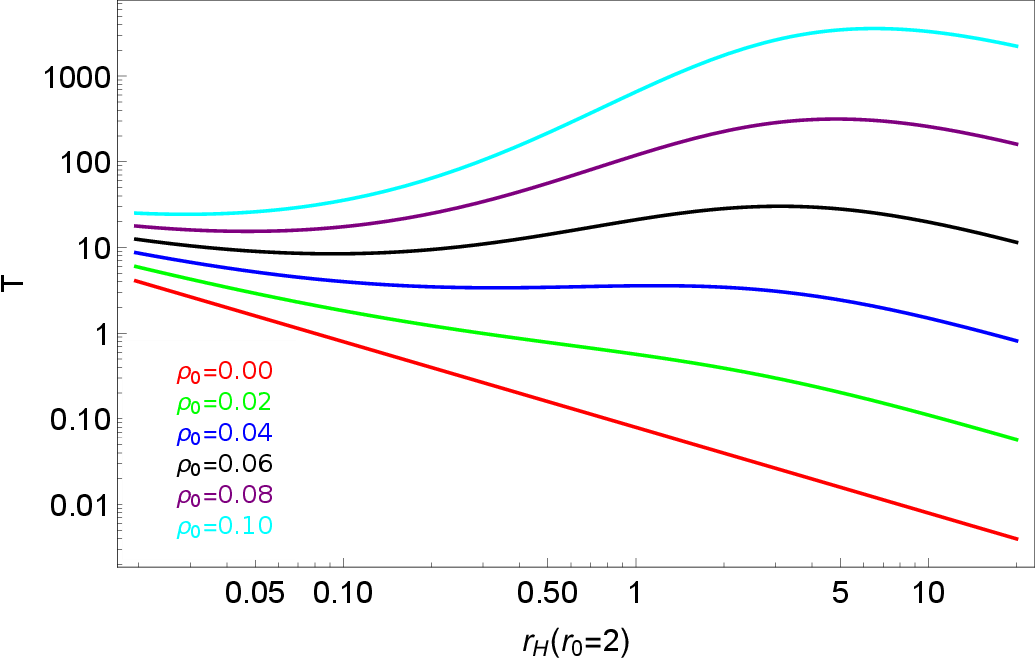}
    \includegraphics[scale=0.35]{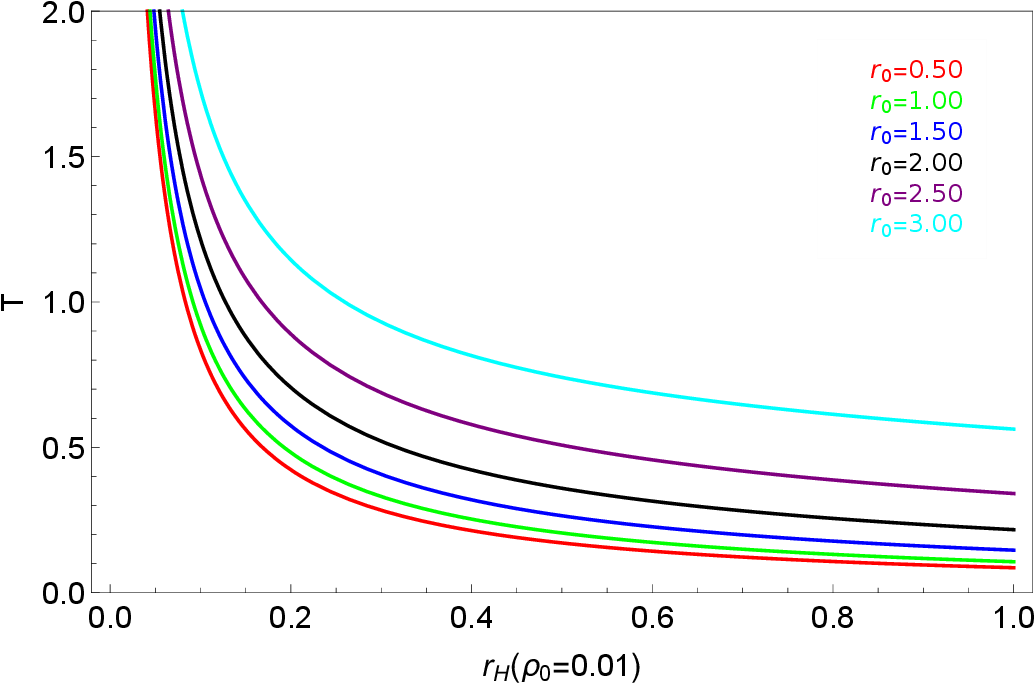}
    \caption{Profile of temperature  as function of $r_H$ for various $\{\rho_0,r_0\}$. The left panel is displayed in a log-log scales. } 
  \label{fig:T}
\end{figure}

The surface gravity of the horizon $\kappa$ defines the Hawking temperature as follows, $T= \frac{\kappa}{2\pi} = \frac{f'+F'}{4\pi}\vert_{r=r_H}$. The temperature reads
\begin{align}
T &= e^{\frac{32}{3}\pi\rho_0r_0^2\left(1+\frac{r_0}{r_H}\right)^{-1/2}}\left[\frac{1}{4\pi r_H}+\frac{4}{3}\frac{\rho_0 r_0^3}{\sqrt{r_H(r_0+r_H)^3}}\right].
\end{align}
Note that if $\rho_0$ is set to zero, we recover the Schwarzschild black hole's Hawking temperature $T=\frac{1}{8\pi M}$. Moreover, at large $r_H$, the temperature decreases as $\frac{1}{r_H}$ and asymptotically approaches zero. Fig.~\ref{fig:T} shows the black hole's temperature $T$ with respect to $r_H$ for various combinations of $\{\rho_0,r_0\}$. The left (right) panel depicts the behavior of $T$ with fixed $r_0=2$ ($\rho_0=0.01$). In the Schwarzschild case ($\rho_0=0$, red curve), the temperature decreases monotonically with $r_H$. Thus, the slope is always negative. For sufficiently large dark matter density $\rho_0$, the temperature profiles have interesting character. At first, the temperature develops to a minimum at a certain $r_H$. As $r_H$ increases, the temperature reaches its maximum before decreasing with $r_H$. This clearly indicates at least three branches of black hole, i.e., the small black hole with negative $\frac{\partial T}{\partial r_H}$, the intermediate black hole with positive $\frac{\partial T}{\partial r_H}$ and sufficiently large black hole with negative  $\frac{\partial T}{\partial r_H}$. For small $\rho_0$, the temperature profile only decreases with $r_H$. 

The black hole's entropy can be computed directly from \cite{Gohain:2024eer}
\begin{align}
    S &=\int \frac{1}{T}\frac{dM}{dr_H}dr_H=\pi r_H^2 =\frac{A}{4},
\end{align}
where the surface area of the black hole is $A=4\pi r_H^2$. This agrees with the Bekenstein black hole's entropy formula \cite{PhysRevD.7.2333} where the entropy of the black hole is proportional to the black hole's surface area. In addition, the dark matter halo has no contribution to the entropy of the event horizon. 

To investigate the thermal stability of a black hole as a thermodynamic system, the heat capacity must be calculated. A positive heat capacity indicates a stable thermodynamic system, while a negative heat capacity indicates instability \cite{Mann:2025xrb,Singh:2025svv}. Positive heat capacity means that the black hole-dark matter system requires energy to increase its temperature, whereas negative heat capacity indicates that the system cools down as heat is injected. In other words, black hole with negative heat capacity gets hotter as it radiates more heat.

\begin{figure}[H]
    \centering
    \includegraphics[scale=0.35]{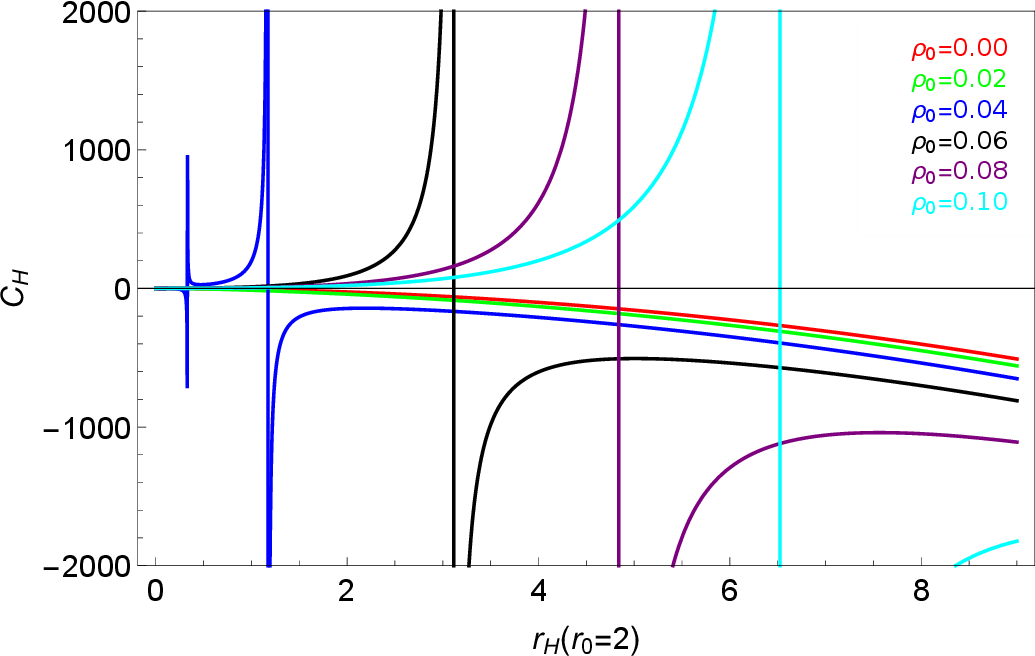}
    \includegraphics[scale=0.35]{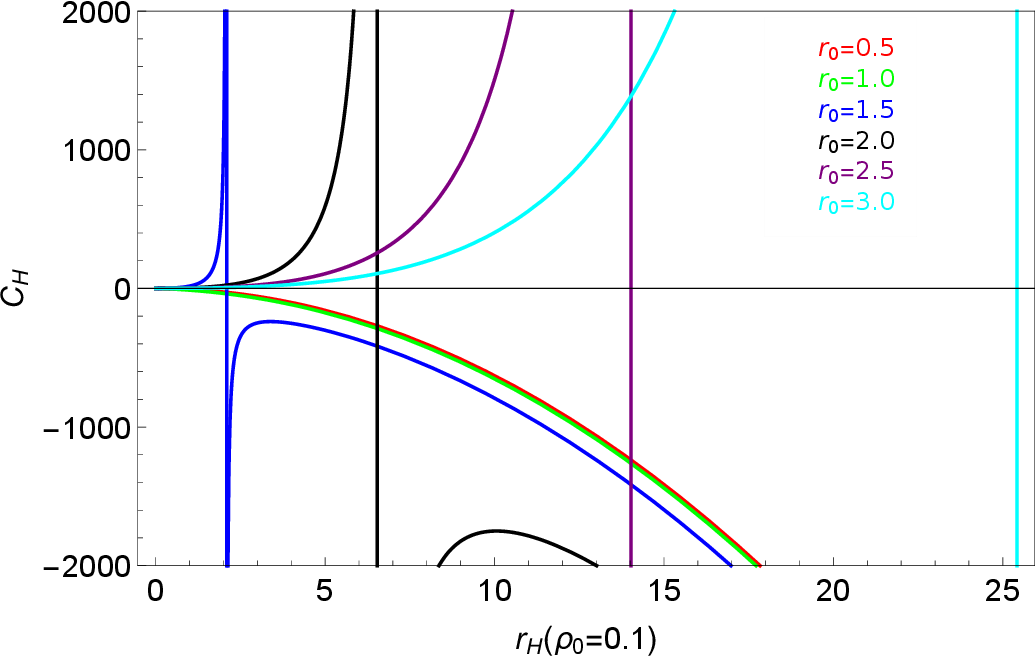}
    \caption{Profile of $C_H$ as function of $r_H$ for various $\{\rho_0,r_0\}$. The case $\rho_0=0$ in the left panel indicates heat capacity of the Schwarzschild black hole.} 
  \label{fig:ch}
\end{figure}

The specific heat capacity is obtained as follows,
\begin{align}
    C_H&=\frac{\partial M}{\partial T}=\frac{\frac{\partial M}{\partial r_H}}{\frac{\partial T}{\partial r_H}}\nonumber\\
    &=-\frac{6 \pi r_H^2 (r_0 + r_H)\left\{3(r_0+r_H)^2 +16 \pi \rho_0 r_0^3 \sqrt{r_H (r_0 + r_H)}\right\} }{9 (r_0 + r_H)^3 - 256 \pi^2\rho_0^2 r_0^6 r_H -24 \pi \rho_0 r_0^3 (r_0-2r_H) \sqrt{r_H (r_0 + r_H) }}. \label{heatcapa}
\end{align}
The heat capacity simply reduces to those of the Schwarzschild black hole $C_H = -2\pi r_H^2$ as $\rho_0 \to 0$ \cite{Kachelriess:2017cfe}. The black hole's heat capacity $C_H$ in relation to $r_H$ for different combinations of $\{\rho_0,r_0\}$ is displayed in Fig.~\ref{fig:ch}. The behavior of $T$ with fixed $r_0=2$ ($\rho_0=0.1$) is seen in the left (right) panel. The behavior of the heat capacity $C_H$ is directly related to the slope of the temperature $T(r_H)$ function. 

It is important to note that the $C_H$ curves break at the extrema of $T(r_H)$, i.e., when $\frac{\partial T}{\partial r_H}=0$. It is clear that the stable (unstable) branch with $C_H > 0 (< 0)$ corresponds to the portion of the temperature function with a positive (negative) slope. Black hole + dark matter configurations without an inflection point, or with a tiny $\rho_0$, do not have a thermodynamically stable branch. In the Schwarzschild limit, temperature decreases monotonically with $r_H$, resulting in a negative heat capacity throughout the domain, highlighting the instability of asymptotically flat black holes in the classical ensemble \cite{Benkrane:2025ukw}.

When dark matter is added, the structure of $T(r_H)$ changes and regions with a positive slope appear, corresponding to domains with a positive heat capacity, indicating a thermodynamically stable branch. Interestingly, when $\rho_0=0.02$ (left panel) and $r_0=1.0$, the heat capacity profile remains similar to that of the Schwarzschild case, indicating that these configurations are thermodynamically unstable. At sufficiently large $\rho_0,r_0$, the heat capacity structure becomes more non-trivial. It is observed that for a fixed $\{\rho_0,r_0\}$, there are two transition points where $C_H$ diverges. The divergence of heat capacity suggests an existence of second-order phase transition \cite{Theor5,roychowdhury2014,Rostami:2020gqq} . The radius at which the transitions occur can be obtained by setting the denominator of \eqref{heatcapa} to zero. For instance, for $\rho_0=0.04$ and $r_0=2$, we find that the transitions locate at $r_H=0.33$ and $r_H=1.17$. Although it cannot be seen easily from the plots in the left panel, we also find two second-order phase transitions for $\rho_0=0.06,0,08,0.10$ (black, purple, cyan). In addition, as $\rho_0(r_0)$ gets larger, the two transition points also grows apart. 

The transition from the unstable to stable at smaller $r_H$ then shift to the unstable again at large $r_H$ implies that sufficiently small and large black holes with the Dehnen dark matter profile are thermodynamically unstable. Transitions from thermodynamically unstable to stable regimes and vice versa, occur at the temperature function's extreme points. Increasing the core dark matter density broadens the range of $r_H$ for which the heat capacity is positive, extending the stable phase of the black hole-dark matter system. In contrast to Schwarzschild black hole case, this shows how the surrounding Dehnen $\left(1,4,\frac{3}{2}\right)$ dark matter halo can act as a stabilizing medium.

To assess the overall stability of the black hole with dark matter system, we study the Gibbs free energy $G$ in our model. A positive $G$ indicates instability in the thermodynamic system, whereas a negative value shows stability. The Gibbs free energy is computed as,
\begin{align}
    G&=M-TS, \nonumber\\
&=e^{\frac{32}{3}\pi\rho_0r_0^2\left(1+\frac{r_0}{r_H}\right)^{-1/2}}r_H\left[\frac{1}{4}-\frac{4}{3}\pi\rho_0 r_0^3\sqrt{\frac{r_H}{(r_H+r_0)^3}}\right].
\end{align}

\begin{figure}[H]
    \centering
    \includegraphics[scale=0.35]{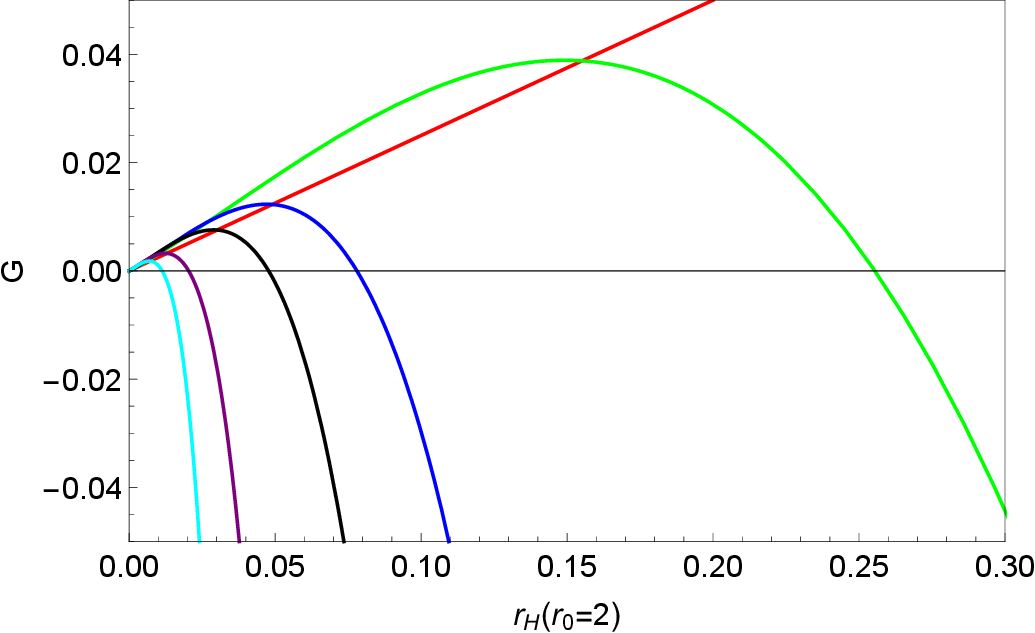}
    \includegraphics[scale=0.35]{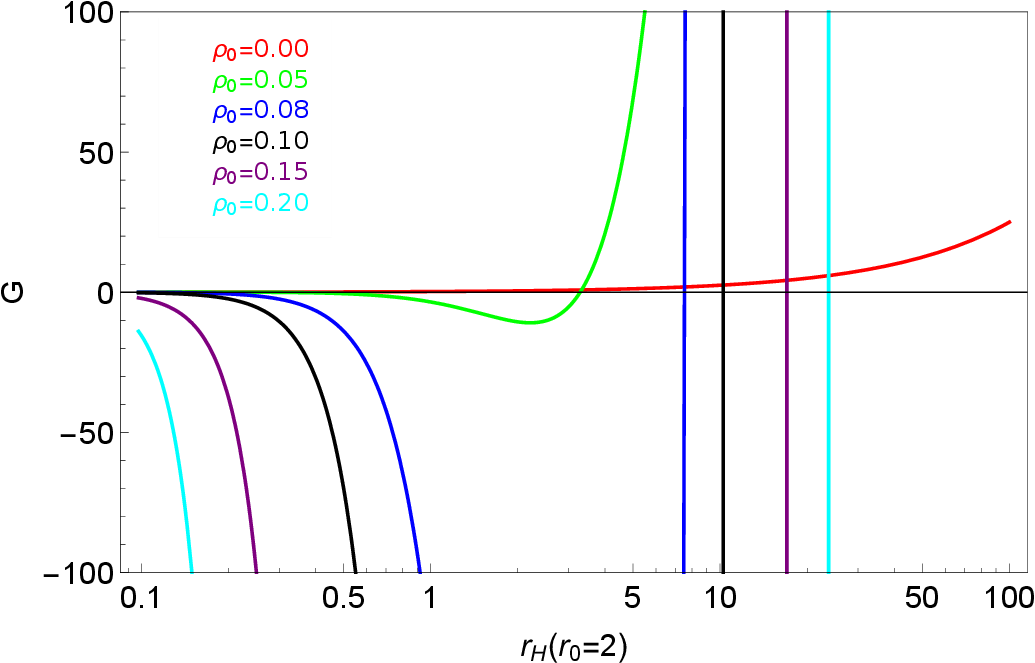}\\
    \includegraphics[scale=0.35]{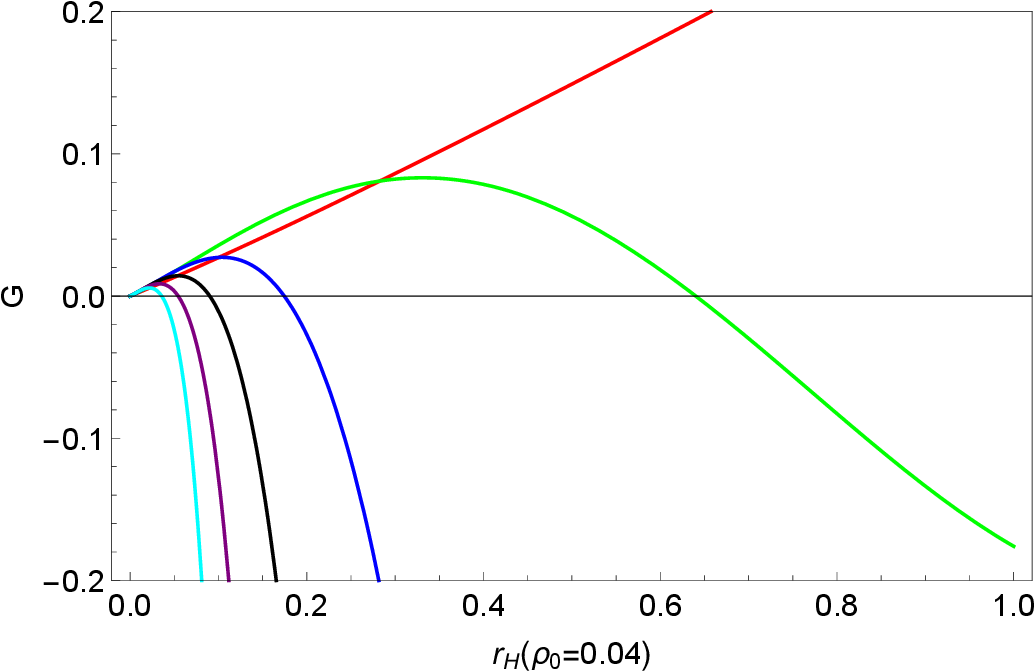}
    \includegraphics[scale=0.35]{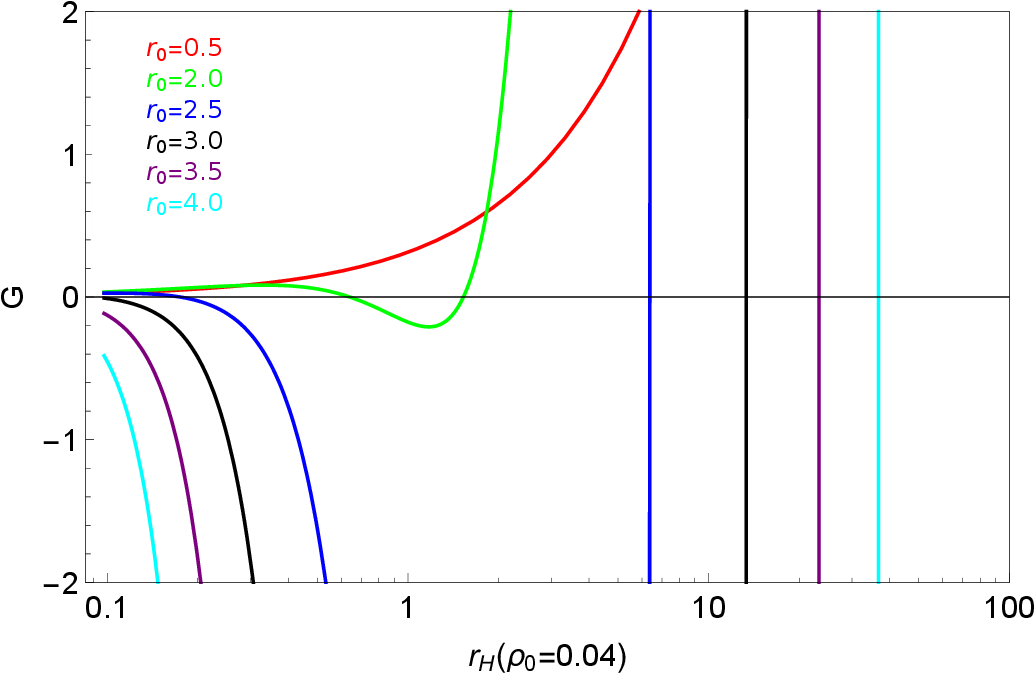}\\
    \caption{Profile of $G$ as function of $r_H$ for various $\{\rho_0,r_0\}$. The left panel depicts the behavior of $G$ at small $r_H$ while the right panel illustrates the behavior of $G$ at a broader range of $r_H$.} 
  \label{G}
\end{figure}

\begin{figure}[H]
    \centering
    \includegraphics[scale=0.5]{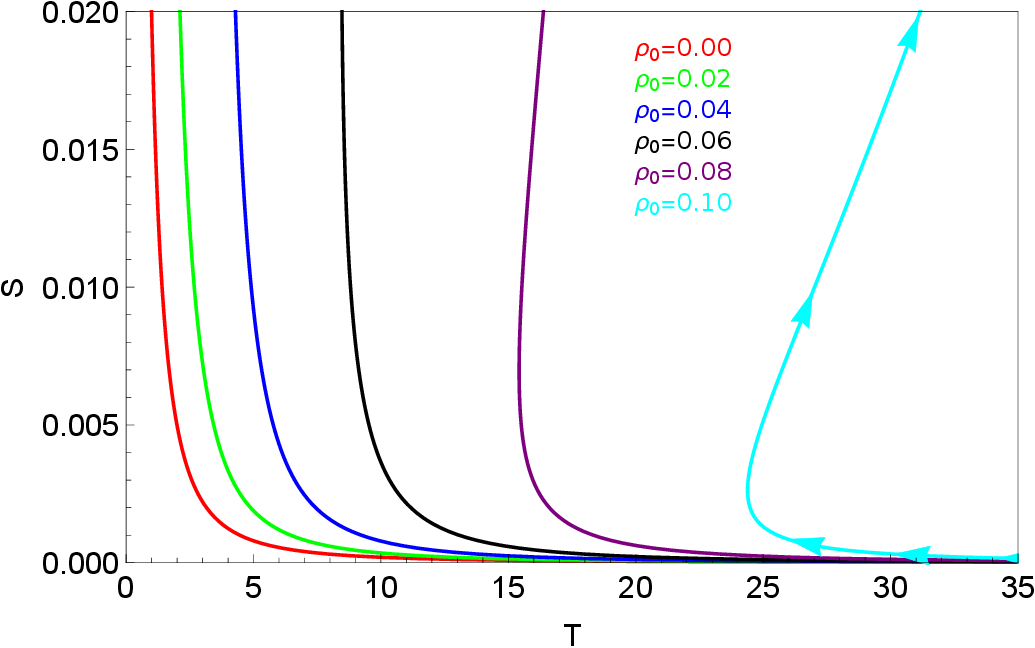}
    \includegraphics[scale=0.5]{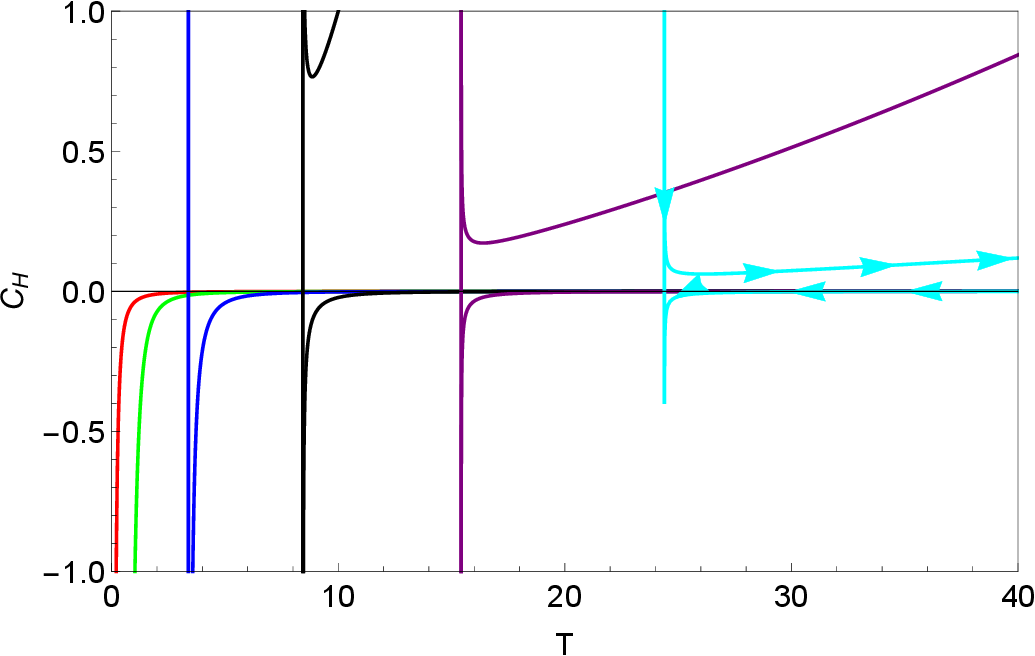}
     \includegraphics[scale=0.5]{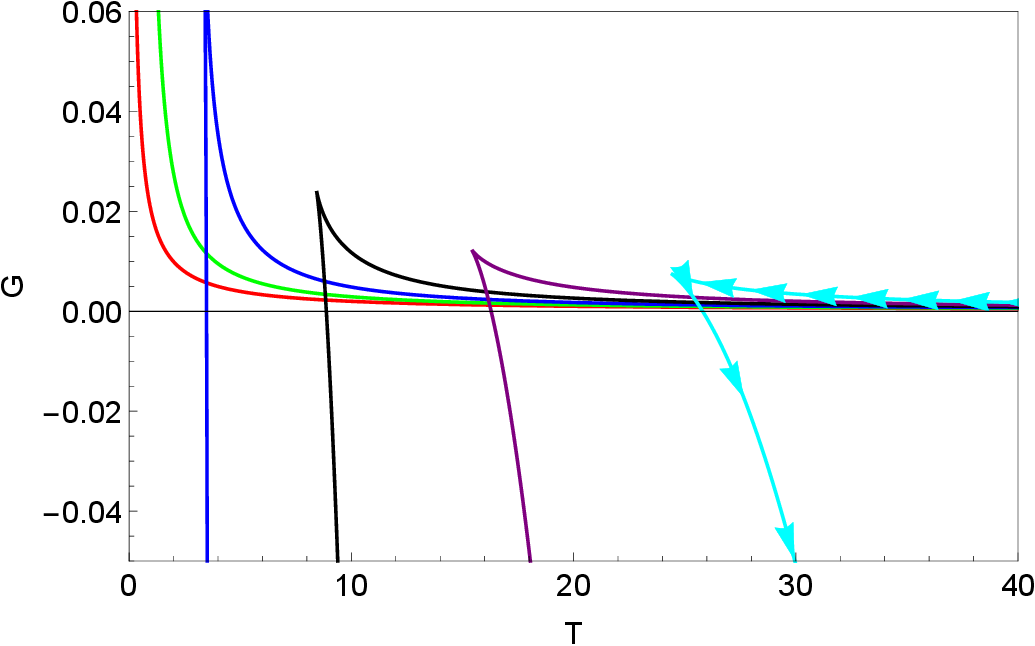}
    \caption{Profile of  $S(T),C_H(T)$ and $G(T)$ for fixed $r_0=2$ and various $\rho_0.$ The arrow indicates the direction of increasing $r_H.$ } 
  \label{Trans}
\end{figure}

Figure~\ref{G} shows the black hole's Gibbs free energy $G$ as a function of $r_H$ for various combinations of $\{\rho_0,r_0\}$. The top (lower) graph shows how $G$ behaves with fixed $r_0=2$ ($\rho_0=0.04$). Our findings indicate that the negative region of the Gibbs free energy increases with higher core density and halo radius. Both graphs demonstrate that the presence of a dark matter halo, in general, increases the black hole's global stability. Later, the Gibbs free energy turns positive at greater $r_H$ indicating that black holes are thermodynamically unstable at large $r_H$.
Figure~\ref{Trans} shows the entropy, heat capacity and Gibbs free energy as functions of temperature with fixed $r_0=2$. In these plots, the arrows mark direction of increasing horizon radius $r_H$. The smaller black holes (small $r_H$) corresponding to lower entropies and higher temperatures, while larger black holes (large $r_H$) occupy the opposite end. Importantly, the entropy remains continuous across the entire temperature range. There is an inflection point at critical temperature for each curve with large enough $\rho_0$ where $\frac{dS}{dr_H}$ changes sign.

The heat capacity $C_H(T)$, however, displays a remarkably different behavior. For small black holes, the heat capacity is negative and as $r_H$ increases, the system encounters divergences in $C_H(T)$, marking second-order phase transitions from small to large black hole. For large black hole regime, the heat capacity is positive, indicating the existence of locally stable black hole branches supported by the halo. More investigation has revealed that for $\rho_0=0.02$ case, the heat capacity remains negative. In contrast, for the $\rho_0>0.02$ cases, we find that as temperature increases another second-order phase transition occurs which marks the transition from positive heat capacity to negative one.

The Gibbs free energy $G(T)$ of small black holes (at high temperatures) exhibit positive free energy and are globally unstable, while large black holes with positive $C_H$ show negative $G(T)$, identifying them as thermodynamically favored states. Moreover, at the critical points where
\begin{equation}
  \frac{\partial G(T)}{\partial T}=  \frac{\partial^2 G(T)}{\partial T^2}=0.
\end{equation}
We observe the following distinct properties: The Gibbs function $G(T)$ and its slope $\frac{\partial G(T)}{\partial T}$ are continuous, whereas $\frac{\partial^2 G(T)}{\partial T^2}$ is not. According to Ehrenfest's classification, it clearly represents a second-order phase transition. Also note that a phase transition occurs if  $\rho_0$ is sufficiently large.

Taken together, Fig.~\ref{Trans} demonstrate that the Dehnen $\left(1,4,\frac{3}{2}\right)$ dark matter halo introduces a stable thermodynamically phase of static spherically symmetric black hole, enriching the phase structure and highlighting how dark matter can act as a thermodynamically stabilizing medium. 

\section{Null Geodesic and Black Hole's Shadow}\label{sec:null}
Null geodesics are the paths followed by light in curved spacetime. Around a black hole, these trajectories determine which photons escape to infinity and which are captured. The boundary between these two outcomes forms the photon sphere and when projected onto the observer’s sky, it defines the black hole’s shadow—a dark silhouette surrounded by a bright emission ring \cite{grenzebach2016shadow}.

Now, let us consider photon trajectory around the Schwarzschild black hole immersed in the Dehnen $\left(1,4,\frac{3}{2}\right)$ dark matter halo. We start by writing the Lagrangian of a massless particle,
\begin{equation}
 \mathcal{L}(x,\dot{x})=\frac{1}{2}g_{ab}\dot{x}^a\dot{x}^b=0,
\end{equation}
where $g_{ab}$ is the metric tensor components given by \eqref{metric} and $\dot{x}^a=dx^{a}(\lambda)/d\lambda$ where $\lambda$ is the curve parameter. Substituting the metric tensor components into the Lagrangian, it is straightforward to obtain the following expression,
\begin{equation}
  \mathcal{L}(r,\theta,\dot{t},\dot{r},\dot{\theta},\dot{\phi}) =\frac{1}{2}\left[-h(r)\dot{t}^2+\frac{\dot{r}^2}{h(r)}+r^2\dot\theta^2+r^2\sin^2\theta \dot\phi^2\right], \label{Lagrangian}
\end{equation}
where $h(r)$ can be found in \eqref{metric}. There are two constants of motion, i.e., in the temporal and the azimuthal directions. This is because the Lagrangian \eqref{Lagrangian} is not an explicit function of $t$ and $\phi$. These constants are
\begin{gather}
    p_t=\frac{\partial \mathcal{L}}{\partial \dot{t}}=-h(r)\dot{t}=-E, \label{4} \\
    p_\phi=\frac{\partial \mathcal{L}}{\partial \dot{\phi}}=r^2\dot{\phi} \sin^2\theta=L\label{5},
\end{gather}
where $E$ and $L$ denote the photon's energy and angular momentum, respectively.

In the equatorial plane $\theta=\frac{\pi}{2}$, the Lagrangian \eqref{Lagrangian} is further simplified,
\begin{align}
  \mathcal{L}(r,\dot{t},\dot{r},\dot{\phi}) &=\frac{1}{2}\left[-h(r)\dot{t}^2+\frac{\dot{r}^2}{h(r)}+r^2\dot\phi^2\right]\nonumber \\&=\frac{1}{2}\left[-\frac{E^2}{h(r)}+\frac{\dot{r}^2}{h(r)}+\frac{L^2}{r^2}\right]=0.  \label{radialnull}
\end{align}
This implies the following radial equation
\begin{align}
    \dot r^2+V_{eff}=\frac{1}{b^2},  \label{radialeq1} 
\end{align}
where we have defined $\bar{\lambda} \equiv \lambda L$ and the impact parameter $b^2=L^2/E^2$. The first and second terms on the left-hand side represent the radial kinetic energy and effective potential $ V_{eff}$, Note that, from this point onward $\dot{r}=dr/d\bar{\lambda}$. The explicit form of the effective potential is 
\begin{align}
    V_{eff}&=\frac{h}{r^2}=\frac{1}{r^2}\left[e^{\frac{32}{3}\pi\rho_0r_0^2\left(1+\frac{r_0}{r}\right)^ {-\frac{1}{2}}}-\frac{r_s}{r}\right].  
\end{align}
A maximum of the effective potential signifies the location of photon circular orbits. These circular orbits are, in fact, unstable. This is because a slight perturbation may cause the photon moving away from the peak. The radius of the photon sphere $r_{ph}$ can be computed from the following conditions
\begin{align}
    \frac{d}{dr}V_{eff}(r_{ph}) &=0,~~~~~~V_{eff}(r_{ph}) = \frac{1}{b_c^2}, \label{rpheq1}
\end{align}
where $b_c$ is the critical impact parameter. The first condition yields 
\begin{align}
    \left(\frac{h'}{r^2} - \frac{2h}{r^3}\right)\bigg\rvert_{r=r_{ph}} &= 0. \label{rpheq}
\end{align}
Since at the event horizon $r=r_H$, the second term above vanishes, while the first term is generally non-zero. This implies that $r_H \neq r_{ph}$. Unfortunately, the photon sphere radius cannot be expressed explicitly due to the complication in \eqref{rpheq}. However, it is possible to show that for a small $\rho_0$, The radius of the photon sphere can be determined from 
 \begin{align}
     0 &= \frac{6-2r_{ph}}{r_{ph}^4} - \frac{16\pi r_0^2}{3r_{ph}^4}\left( 4r_{ph}+3r_0                  \right)\left(\frac{r_{ph}}{r_{ph}+r_0}\right)^{3/2}\rho_0 + \mathcal{O}(\rho_0)^2. \label{photonsphereeq}
 \end{align}
At the zero order, the photon sphere radius simply reduces to those of the Schwarzschild case $r=r_{ph}=3$. On the other hand, the critical impact parameter can be obtained anallytically
\begin{align}
    b_c &= \pm \frac{r_{ph}^{3/2}}{\sqrt{r_{ph}e^{32\pi r_0^2 \rho_0\left(\frac{r_{ph}}{r_{ph}+r_0}\right)^{1/2}}-2}}.
\end{align}
In the Schwarzschild limit ($\rho_0=0,r_{ph}=3)$, the critical impact parameter above becomes $b_c=3\sqrt{3}$. 

At two asymptotic region, the effective potential and its derivative take the form
\begin{align}
V_{eff}(r\to 0) &= -\frac{2}{r^3} + \mathcal{O}\left(1/r^2\right),~~~V'_{eff}(r\to 0)=\frac{6}{r^4}+ \mathcal{O}\left(1/r^3\right), \nonumber \\
V_{eff}(r\to\infty) &= \frac{e^{32\pi r_0^2\rho_0/3}}{r^2} + \mathcal{O}\left(1/r^3\right),~~~V'_{eff}(r\to\infty)=-2\frac{e^{32\pi r_0^2\rho_0/3}}{r^3}+ \mathcal{O}\left(1/r^4\right).
\end{align}
Given that $V_{eff}(r)$ starts negative with an increasing slope as $r\to 0$ and becomes positive with a decreasing slope as $r\to \infty$ continuity implies the existence of at least one point in between where the derivative of $V_{eff}(r)$ vanishes, i.e., a maximum. The photon sphere is guaranteed if the maximum occurs outside the event horizon. 

\begin{figure}[h]
    \centering
    \includegraphics[scale=0.35]{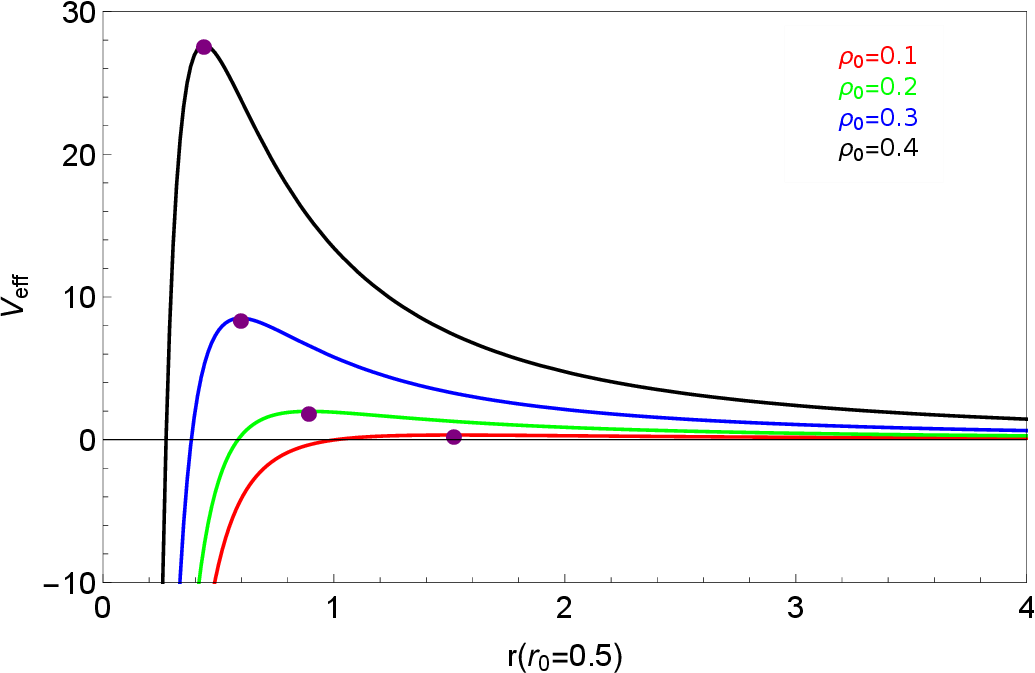}
    \includegraphics[scale=0.35]{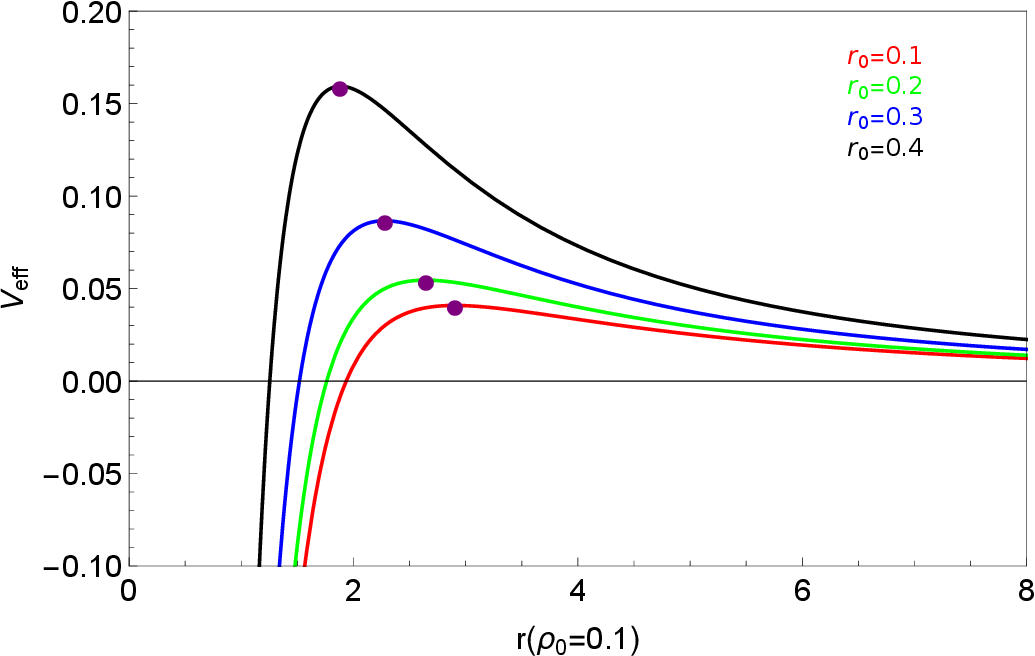}
    \caption{Profile of $V_{eff}$ for selected combinations of black hole parameter. The purple dots mark the location of $V_{eff}(r_{ph})=V_{max}$.} \label{fig:veff}
\end{figure}

Figure~\ref{fig:veff} shows a graph of the effective potential behavior for different black hole parameters combinations with respect to $r$. It is observed that as $\rho_0$ and $r_0$ increase, the maximum height of the effective potential ($V_{max}$) increases. This implies that the critical impact parameter $b_c$ decreases with $\rho_0$ and $r_0$. Moreover, the unstable photon circular orbit ($r_{ph}$) becomes smaller while $\rho_0$ and $r_0$ get larger.

\begin{figure}[h]
    \centering
    \includegraphics[scale=0.25]{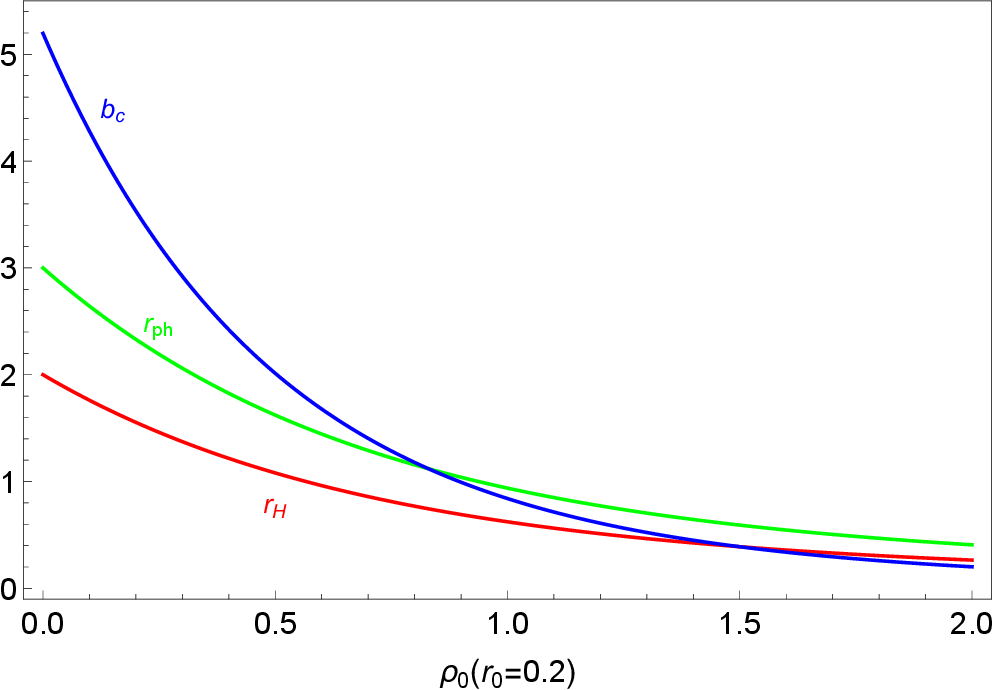}
    \includegraphics[scale=0.25]{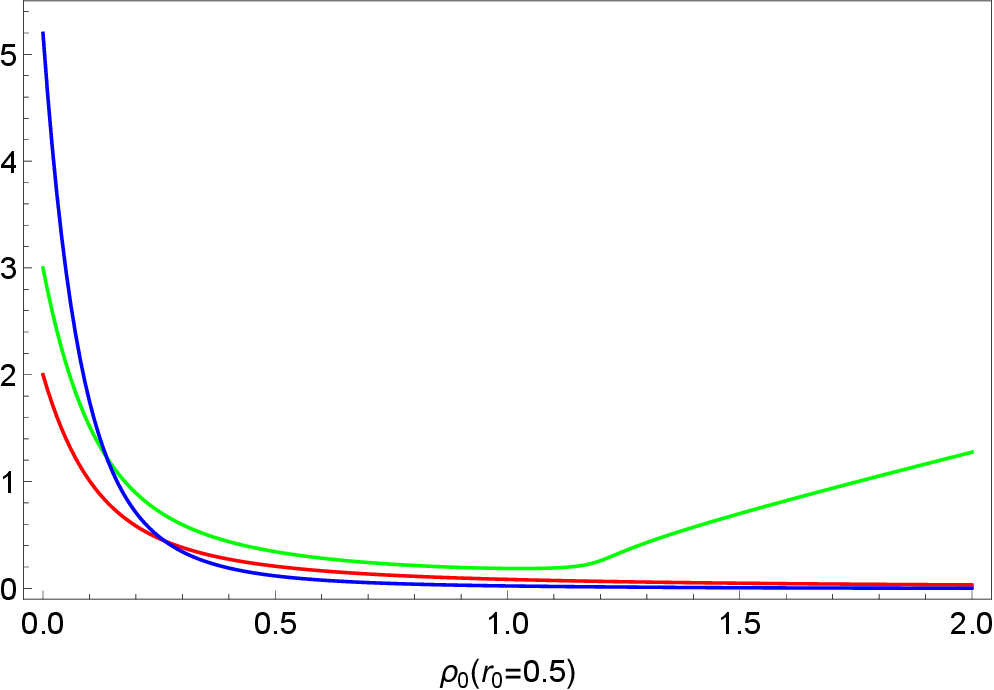}
     \includegraphics[scale=0.25]{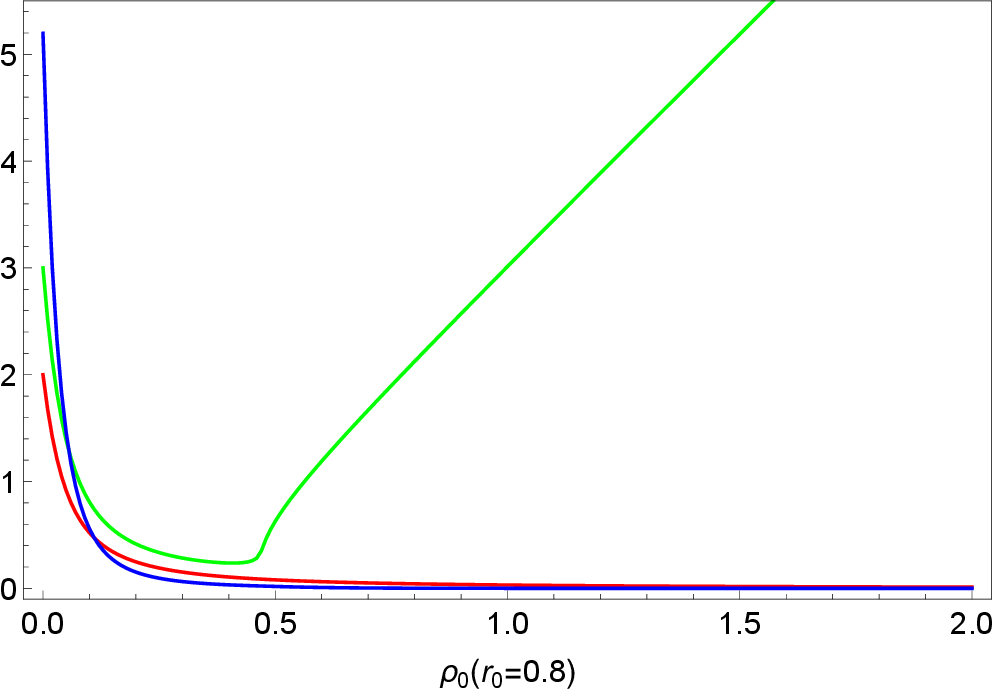}
    \caption{The behavior of $r_H,r_{ph}$ and $b_c$ as function of $\rho_0$ for three specified values of $r_0$.} \label{fig:rhrphbc}
\end{figure}

Additionally, black hole's event horizon ($r_H$), photon sphere radius $(r_{ph})$ and critical impact parameter ($b_c$) are plotted against halo's central density $\rho_0$ as illustrated in Fig.~\ref{fig:rhrphbc}. First of all, at $\rho_0=0$, these parameters resemble those of the Schwarzschild black hole i.e., $r_H=2,r_{ph}=3,b_c=3\sqrt{3}$. At small $r_0=0.2$, we observe a clear monotonic decreasing behavior of these parameters confirming our earlier discussion. As core's radius gets larger, we notice that the photon sphere radius instead of decreasing with $\rho_0$ but increasing after a certain value of $\rho_0$. Moreover, at larger $r_0$, the increasing of $r_{ph}$ against $\rho_0$ occurs at smaller $\rho_0$. The plot also depicts that the event horizon and photon sphere radius never coincide.  

One can consider the case $\rho_0 \gg 1$ in the region with sufficiently large $r_{ph}$ compared to $r$. Under this condition, \eqref{rpheq} can be approximated as
\begin{equation}
-2 + \frac{16 \pi r_0^3 \rho_0}{ r_{ph}} = 0 \quad \Rightarrow \quad r_{ph} = 8 \pi r_0^3 \rho_0,
\end{equation}
which shows that $r_{ph}$ scales linearly with $\rho_0$ with slope proportional to $ r_0^3$. 

\subsection{Black Hole's Shadow}
The propagation of photons in the strong gravitational field of a black hole exhibits a rich variety of behaviors that underlie key observational phenomena. Photon trajectories can be categorized into three distinct classes: those captured by the event horizon, those temporarily stay in unstable circular orbits and those deflected to infinity after partial revolutions. Only the latter two classes are accessible to a distant observer, with the unstable circular orbit— known as the photon sphere—defining the critical boundary between captured and escaped . The properties of this orbit govern the structure of black hole shadows. To a distant observers, black holes are illuminated by a bright photon ring. The inner edge of it defines black hole's shadow.

Additionally, the circular orbit of photons is fundamentally unstable. Small perturbations determine whether the object spirals toward or escapes the black hole. As a result, backward ray tracing of these departing photons from the photon sphere will travel through the intervening space, influenced by dark matter. These photons originate in the photon sphere and travel reaching the observer at the critical impact parameter. Thus, we can observe how the photon sphere and impact parameter connect to the shadow's perceived angular radius and radius.

Using \eqref{rpheq1}, we can write the critical impact parameter $b_c$ as,
\begin{equation}
    b_c^{-1}=\sqrt{\frac{h(r_{ph})}{r^2_{ph}}}. \label{bcrit}
\end{equation}
An observer at spatial infinity can measure the shadow radius radius $R_s$ as \cite{Pantig:2024rmr}, 
\begin{equation}
R_s=b_c \sqrt{h(r\to\infty)} \approx\sqrt{\frac{r^2_{ph}}{h(r_{ph})}}e^{ \frac{16}{3}\pi\rho_0r_0^2}.  \label{Rs}
\end{equation}
In asymptotically flat spacetimes such as the Schwarzschild or Reissner-Nordstr\"om black holes, the redshift function satisfies $h(r \to \infty) \to 1$. Consequently, the shadow radius measured by a distant observer coincides with the critical impact parameter, $R_s = b_c$. In the present case, the Dehnen $\left(1,4,\frac{3}{2}\right)$ halo alters the asymptotic structure of the spacetime, giving $h(r \to \infty) = e^{\tfrac{16}{3}\pi\rho_0 r_0^2}$ which is generally non-unity. The metric therefore approaches a \textit{conformally flat} rather than a truly asymptotically flat geometry. As a result, the propagation of photons from the photon sphere to infinity results in an additional constant redshift factor, as given in \eqref{Rs}. Interestingly, this distinction is due to the effect of the surrounding dark-matter halo on optical geometry.

For special case with $\rho_0 r_0^3\ll 1$, we can perform series expansion to \eqref{photonsphereeq} and obtain the radius of photon sphere, 
\begin{equation}
    r_{ph}\approx\frac{3(3 r_s + 16 \pi\rho_0 r_0^3 )}{2(3 + 32 \pi\rho_0 r_0^2 )},
\end{equation}
yielding,
\begin{equation}
    R_s\approx\frac{\sqrt{3}}{2}\left[3r_s+16\pi \rho_0 r_0^3\left(1-2\frac{r_s}{r_0}\right)\right],
\end{equation}
where the expression $R_s=\frac{3\sqrt{3}}{2}r_s$ is recovered for the case without dark matter, which is obtained in \cite{Perlick:2021aok}.

\begin{figure}[H]
    \centering
    \includegraphics[scale=0.35]{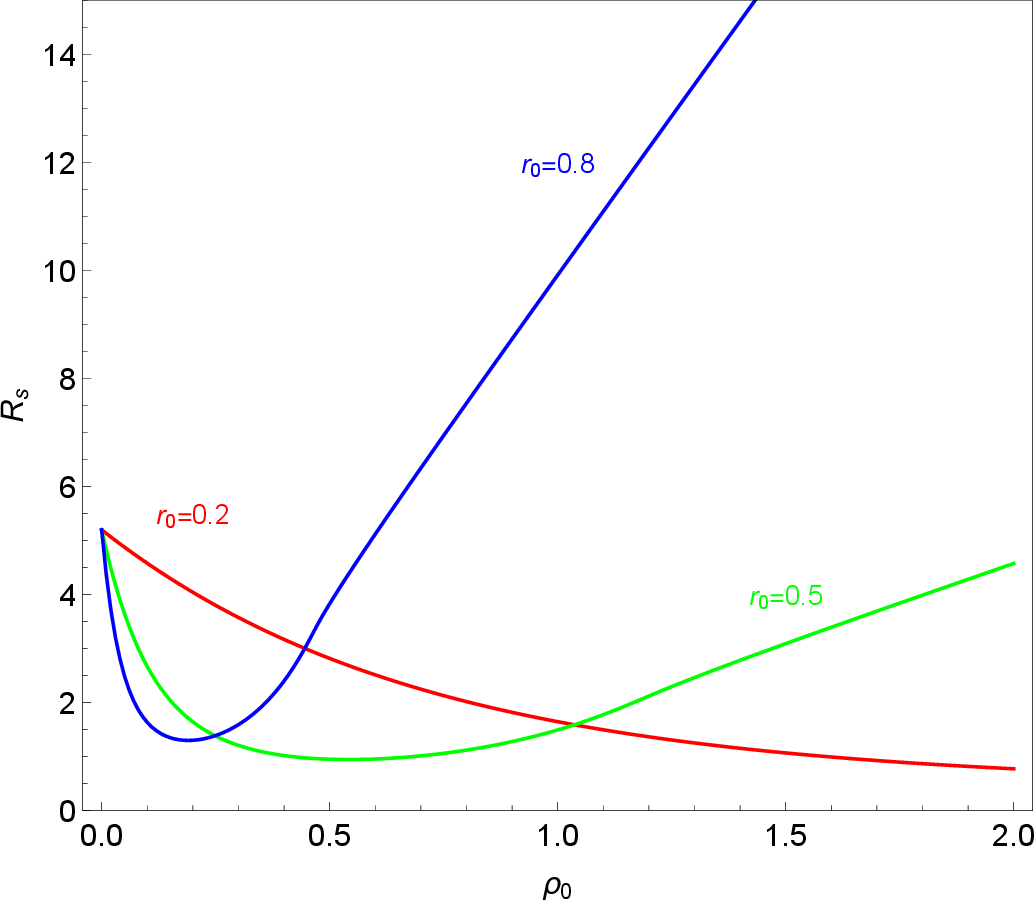}
    \includegraphics[scale=0.35]{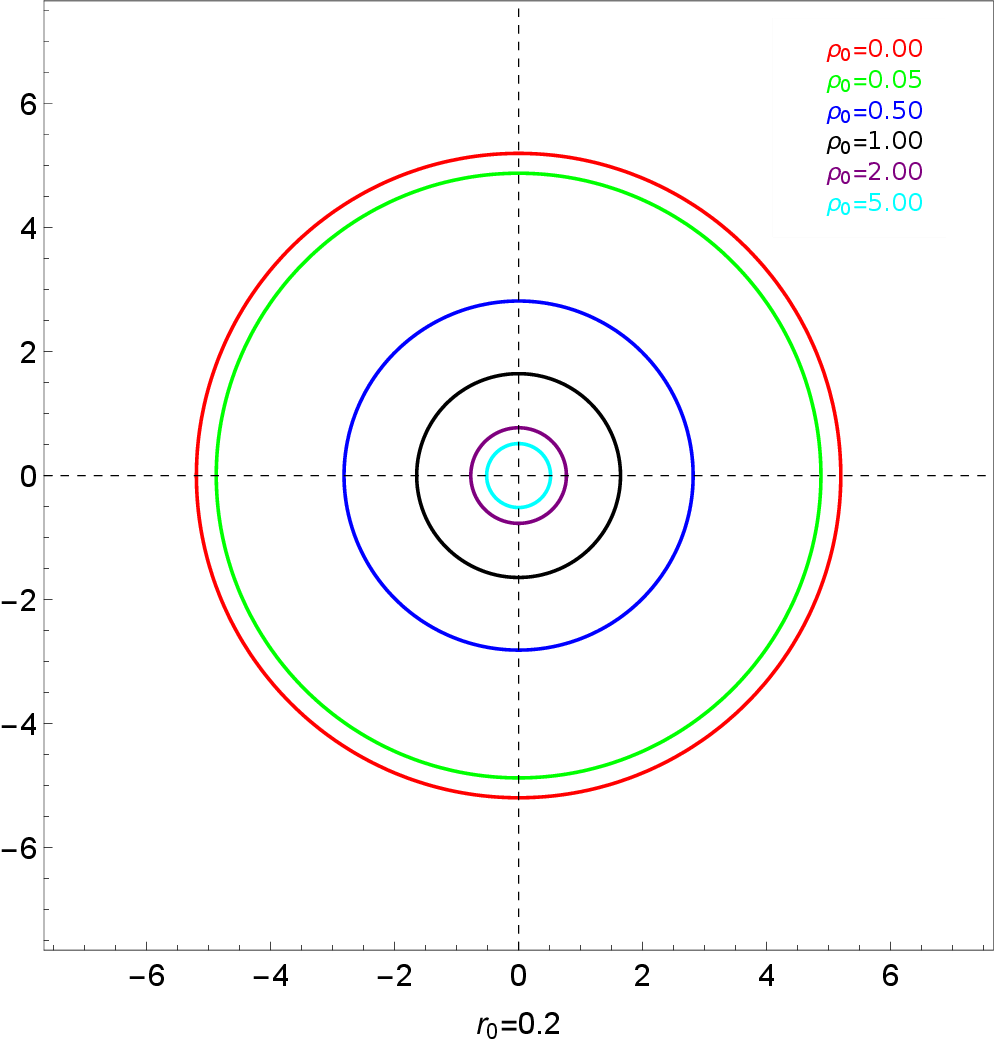}\\
    \includegraphics[scale=0.35]{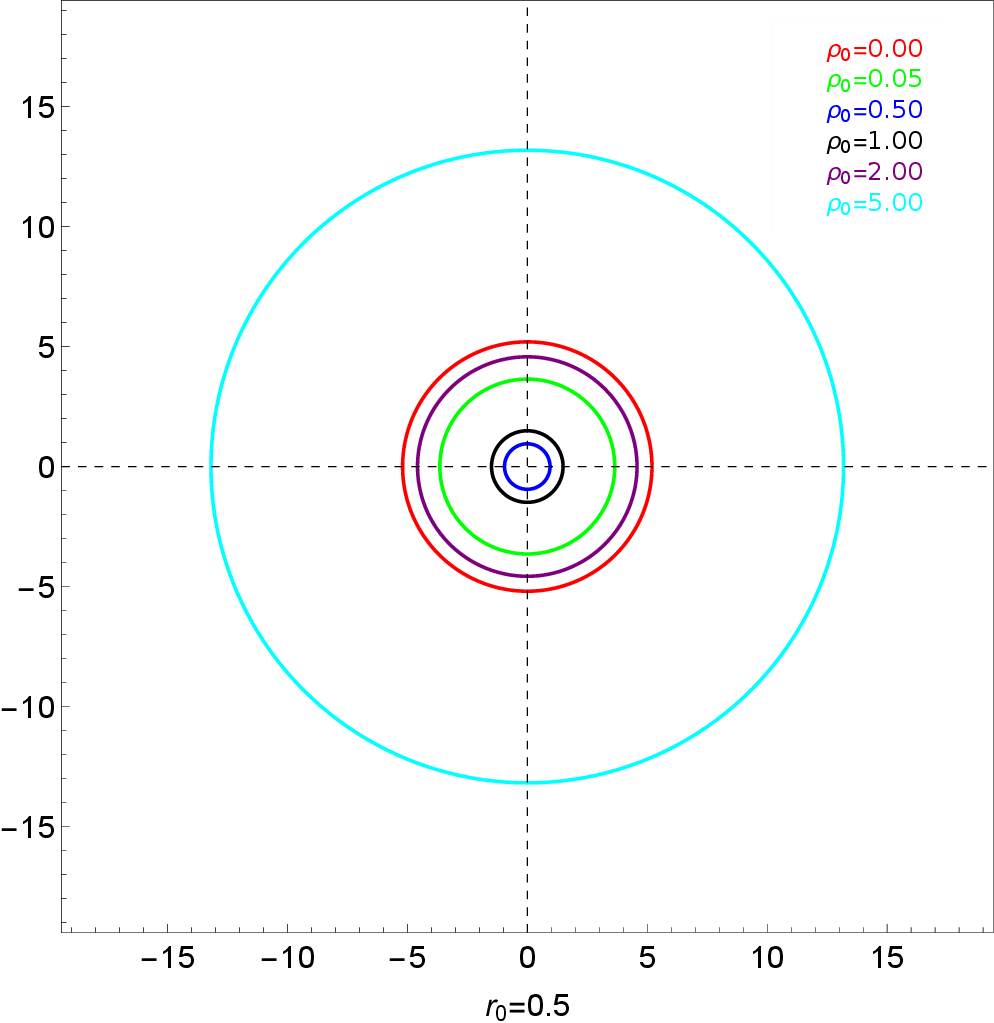}
    \includegraphics[scale=0.35]{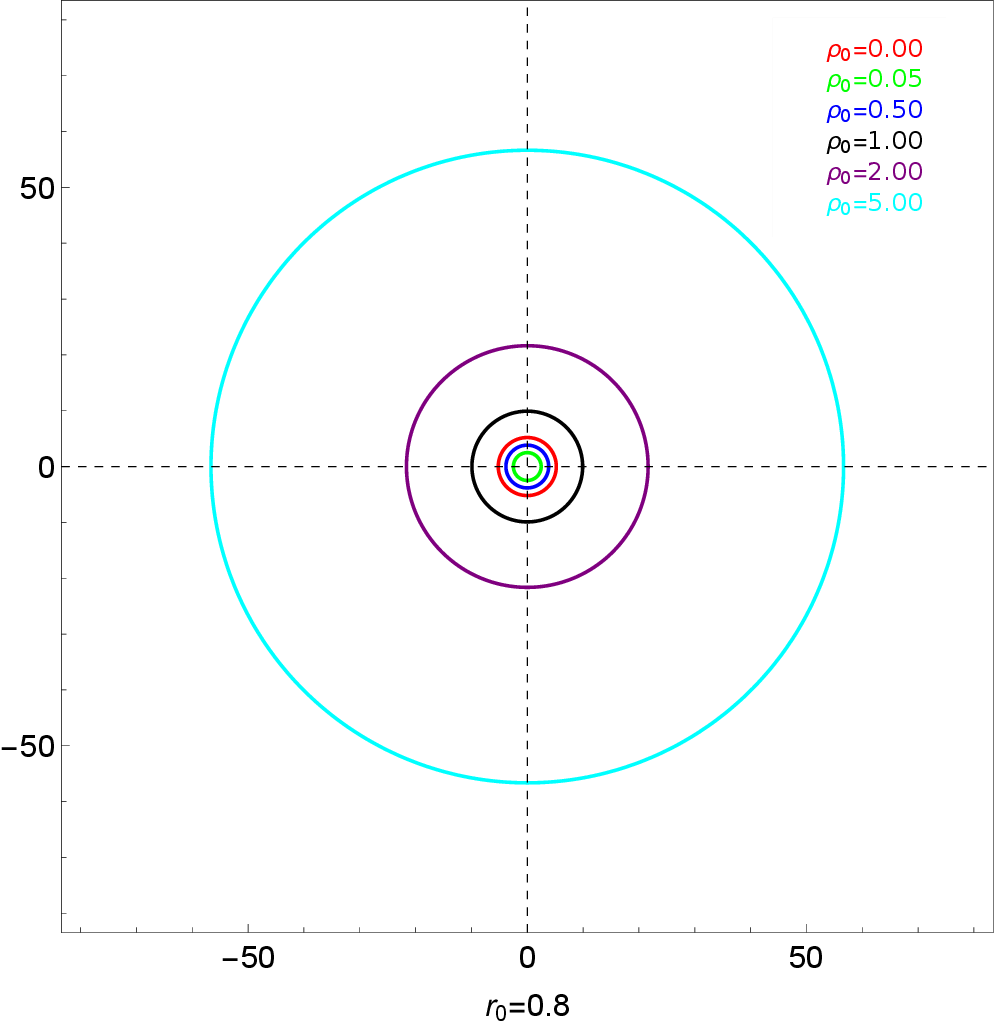}
    \caption{Upper left: Shadow radius $R_s$ as a function of $\rho_0$ for various $r_0=0.2,0.5,0.8.$ Upper right and lower panels: Shadow casts as observed by distance observers. In these figures, we set $r_s=2$. } 
  \label{R}
\end{figure}

Figure \ref{R} depicts shadow casted by black holes as seen from remote observer. The upper left panel displays the shadow radius $R_s$ against halo's central density $\rho_0$ for three distinct value of $r_0$.  At small $r_0$, $R_s$ decreases monotonically with $\rho_0$. As $r_0$ increases, we observe that the shadow radius initially decreases and later increases with $\rho_0$. This implies that at each fixed $r_0$, the radius takes the smallest value at certain value of $\rho_0$. Our numerical investigation reveals that the smallest shadow radii for $r_0=0.2,0.5,0.8$ cases occur at $\rho_0=4.05,0.54$ and $0.19$ respectively.  

Moreover, we display shadow radius for various configurations of $\{\rho_0,r_0\}$. in the upper right, lower left and right panel (for $r_0=0.2,0.5,0.8$ respectively). The red circles in each panel present the shadow radius of the Schwarzschild black hole $(\rho_0=0), R_s=3\sqrt{3}=5.19615$. It is clear that as $\rho_0$ increases, the shadow radius becomes smaller for $r_0=0.2$. At larger $r_0$, the shadow radius behavior turns out to be more complicated. At first, the radius decreases with $\rho_0$ and it reaches the smallest size at $\rho_0=0.5~ (0.05)$ for the lower left (right) panel. After that the radius is larger as $\rho_0$ increases.

\begin{figure}[H]
    \centering
    \includegraphics[scale=0.35]{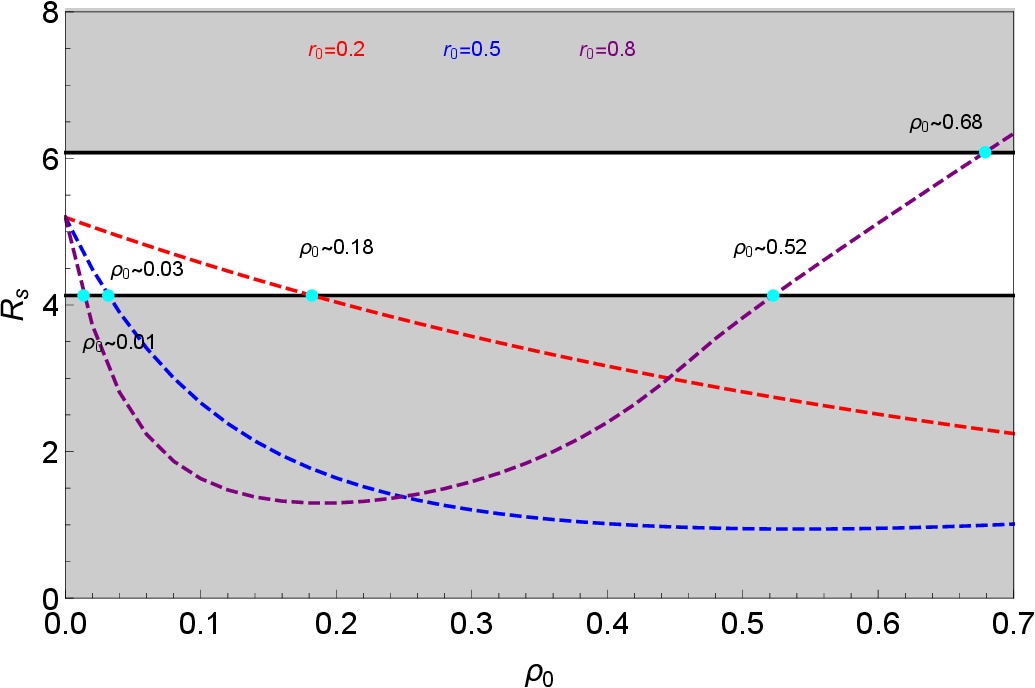}
    \includegraphics[scale=0.35]{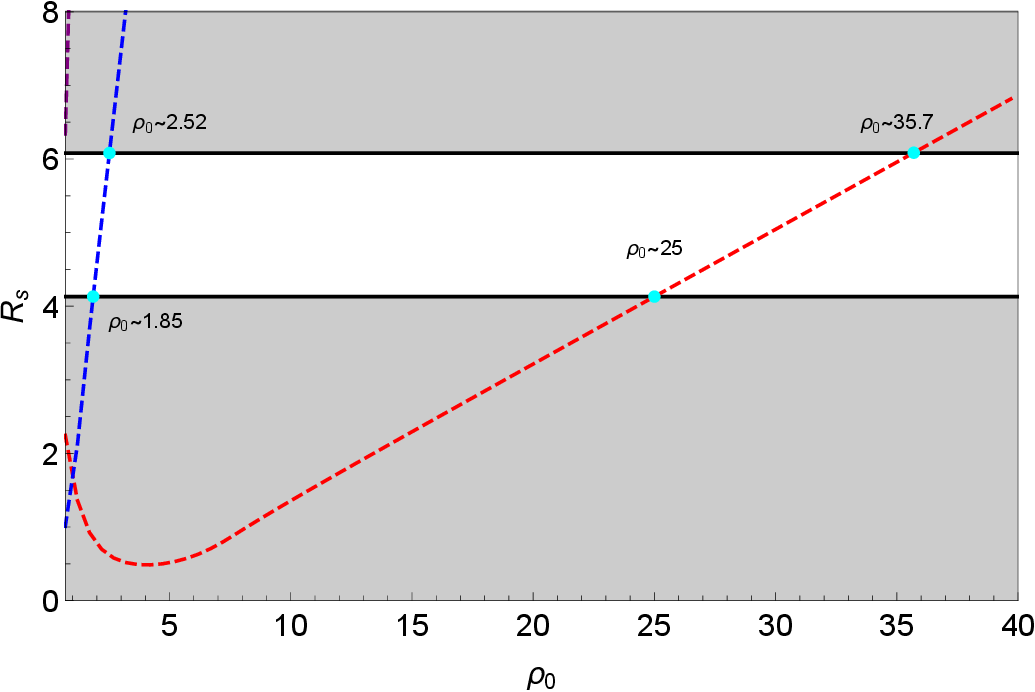}\\
    \includegraphics[scale=0.35]{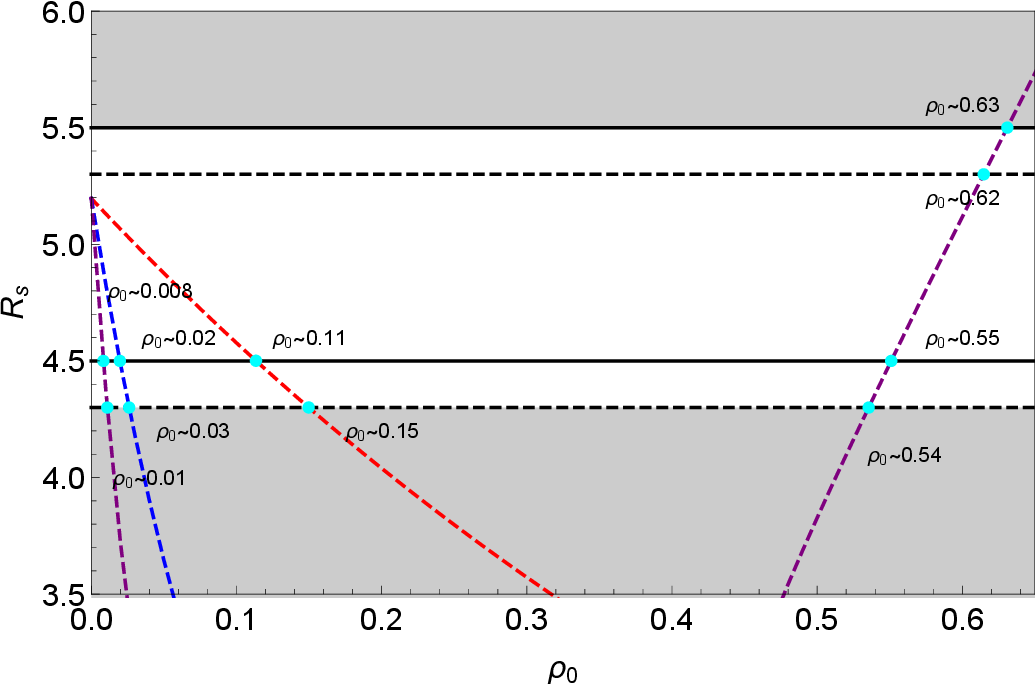}
    \includegraphics[scale=0.35]{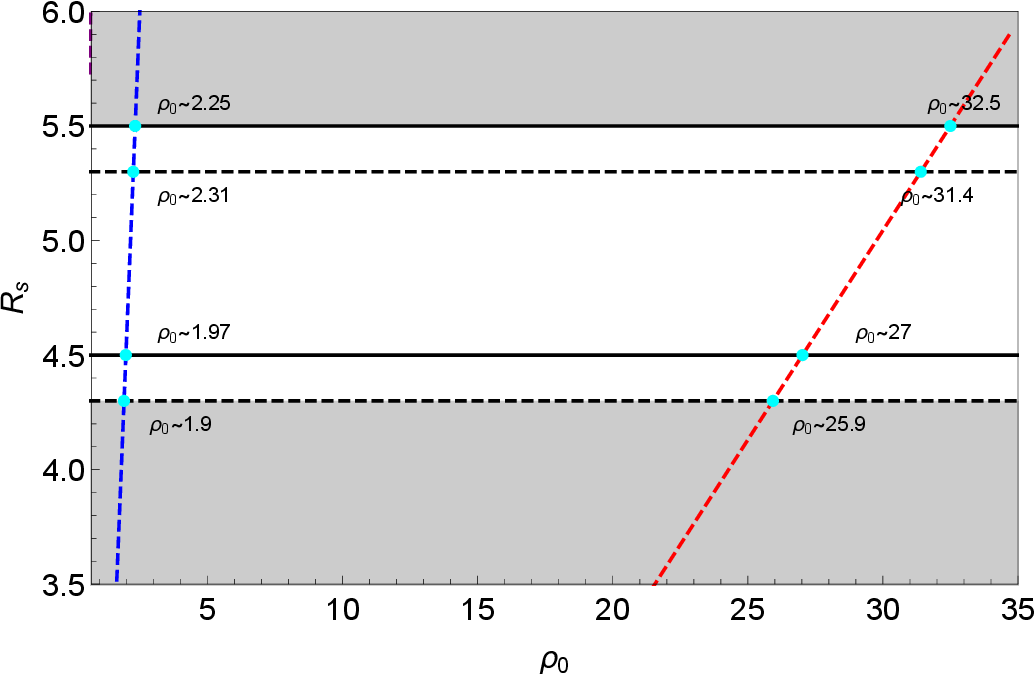}
    \caption{Constraint on shadow Radius $R_s$ expresses in term of $\rho_0$ compared with $M87^\ast$ data (top) and $SgrA^\ast$ data (bottom). There are three distinct profiles of $r_0=0.2,0.5$ and $r_0=0.8$ illustrated by red, blue and purple curved. The cyan dots mark the value $\rho_0$ at the lower and upper bounds of $R_s$. The left panel represents configuration with small $\rho_0$ while the right panel is the same configuration but with larger $\rho_0$. Top: The region where the shadow radius is inconsistent with $M87^\ast$ data is denoted by gray color. Bottom: The allowed region is denoted by white color where the bounds from Keck (VLTI) are marked with black solid (dashed) horizontal lines.} 
  \label{constraintRs}
\end{figure}

By using observational data of $M87^\ast$ from the event horizon telescope (EHT) collaboration \cite{EventHorizonTelescope:2019ggy}, the physical charges of black hole are constrained by using black hole shadow's radius \cite{EventHorizonTelescope:2021dqv}. It is found that the shadow size of $M87^\ast$ must lie within the range
\begin{align}
    4.31 M_{BH} &\leq R_s \leq 6.08 M_{BH}.
\end{align}
For $SgrA^\ast$, the bounds are 
\begin{align}
    4.5 M_{BH} &\leq R_s \leq 5.5 M_{BH}, \\
    4.3 M_{BH} &\leq R_s \leq 5.3 M_{BH},
\end{align}
from the Keck Observatory and the Very Large Telescope Interferometer, respectively \cite{EventHorizonTelescope:2022xqj}. In Fig.~\ref{constraintRs}, we display shadow radius $R_s$ as a function of $\rho_0$ together with the constraints from the $M87^\ast$ and $SrgA^\ast$. In these figures, we fix $r_0=0.2,0.5$ and $0.8$ which are represented by red, blue and purple dashed lines. The gray regions denote the area where black hole's shadow radii are inconsistent with constraints from the $M87^\ast$ and $SrgA^\ast$. There are, in general, two intervals of $\rho_0$ that makes $R_s$ of the Dehnen-black hole are compatible with the observation data depending on the chosen $r_0$.

Finally, it is worth noting that while astrophysical black holes like $SgrA^\ast$ and $M87^\ast$ do rotate, using the constraints from these sources to a non-rotating black hole may lead to unrealistic conclusions. However, it has been demonstrated in \cite{Pantig:2024rmr,Vagnozzi:2022moj} that these sources can nevertheless effectively constrain the parameters of non-rotating spacetime geometry.

\subsection{Weak Gravitational Lensing}
Now we examine the weak deflection angle of particles in a static black hole immersed in a Dehnen-type $\left(1,4,\frac{3}{2}\right)$ dark matter. Weak gravitational lensing (WGL) is a powerful observational technique that enables the measurement of mass distributions in the universe, including galaxies, clusters and large-scale structures. By analyzing the subtle distortions imprinted on the images of background sources, WGL provides a direct and unbiased probe of both luminous and dark matter, making it an essential tool for cosmology and astrophysics \cite{Javed:2022rrs,Jha:2025xjf}. To calculate the weak-field deflection angle, we use the Gauss-Bonnet theorem as applied to the optical metric and Gibbons-Werner method to calculate the weak
deflection angle \cite{Gibbons:2008rj,Waseem:2025yib}. 

The deflection angle $\alpha$ is calculated from the following integral \cite{Mandal:2023eae},
\begin{equation}
    \alpha=\int_{\phi= 0}^{\pi}\int_{r=\frac{b}{ \sin \phi}}^{\infty} K \sqrt{g_{opt}} drd\phi,
\end{equation}
where $K$ and $g_{opt}$ are the Gaussian (3D) curvature and the determinant of the following optical metric,
\begin{equation}
    dt^2=\frac{dr^2}{h^2(r)}+\frac{r^2 d\phi^2}{h(r)},
\end{equation}
yielding,
\begin{equation}
    g_{opt}=\frac{r^2}{h^3(r)},
\end{equation}
and,
\begin{equation}
    K=\frac{1}{2}\left[h(r)\frac{d^2h(r)}{dr^2}-\frac{1}{2}\left(\frac{dh(r) }{dr}\right)^2\right].
\end{equation}
Using \eqref{metric}, we can write the integrand explicitly as a series up to $O(\rho_0 r_0^3)$ as follows,
\begin{equation}
    K \sqrt{g_{opt}} \approx \frac{r_s}{r^2}-\frac{16 \pi \rho_0 r_sr_0^2}{3r^2}+\frac{8\pi\rho_0 r_0^3}{r^3}\left(\frac{2r}{3}+r_s\right).
\end{equation}
This leads to
\begin{equation}
     \alpha\approx \frac{2}{3b}\left[(3 - 16 \pi \rho_0 r_0^2) r_s + \rho_0\pi r_0^3 (16 + 3  \pi b^{-1} r_s) \right]. \label{weakalpha}
\end{equation}
Note that if $\rho_0=r_0=0$, we obtain $\alpha=2 b^{-1} r_s$ which is the deflection angle of an isolated Schwarzschild black hole \cite{Waseem:2025yib}. From \eqref{weakalpha}, the positivity of deflection angle is guaranteed for $0<\rho_0 < \frac{3}{16\pi r_0^2}$. Hence, this yields an upper bound of $\rho_0$ when $r_0$ is fixed. Otherwise, the positive deflection angle depends entirely on the mixing of $b,\rho_0,r_0$. On the other hand, it is possible to obtain $\alpha<0$ once the following conditions are met simultaneously
\begin{align}
    \rho_0 > \frac{3}{8\pi r_0\left(2-r_0\right)},~~~~ b > \frac{3\rho_0\pi^2 r_0^3}{8\pi\left(2-r_0\right)\rho_0 r_0^2-3},~~~~ 0<r_0<2. \label{negalpha}
\end{align}
A negative deflection angle implies that light rays are bent away from the black hole rather than toward it. This behavior can be interpreted as an effective repulsive gravitational influence. Similar repulsive effects have been reported in several previous studies (see \cite{Panpanich:2019mll,Pantig:2022toh,Liu_2024,Molla:2025yoh} and references therein).  In this case, the sign convention for the deflection angle means that counter-clockwise deviations are taken as positive, while clockwise deviations are negative, causing the light rays to miss the intended location of the receiver \cite{Pantig:2022toh,Liu_2024}.

\begin{figure}[H]
    \centering
    \includegraphics[scale=0.35]{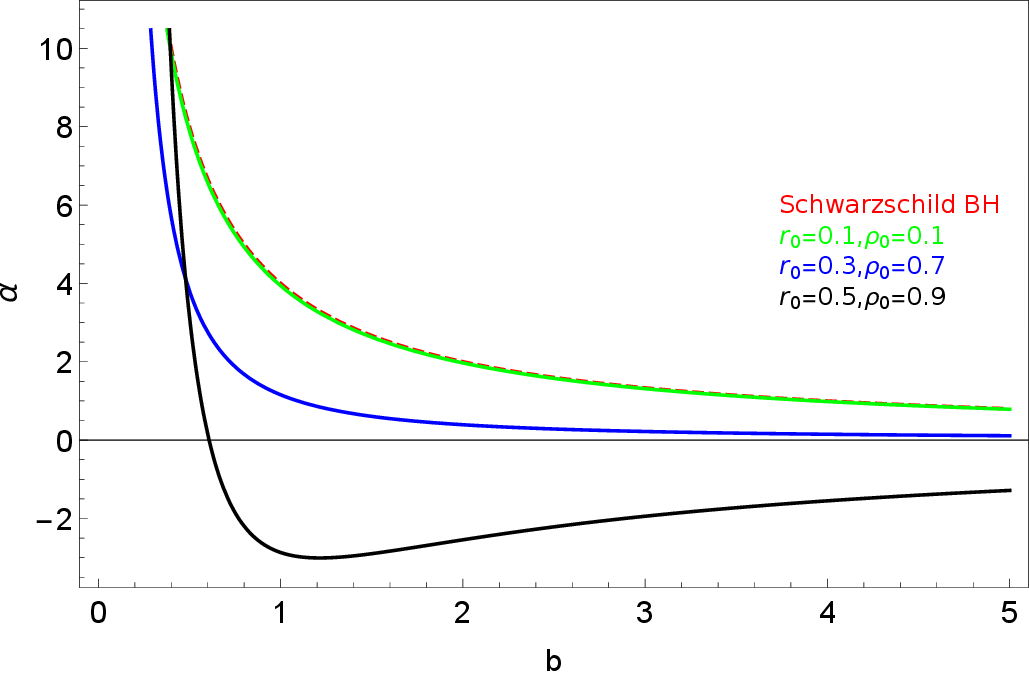}
    \includegraphics[scale=0.35]{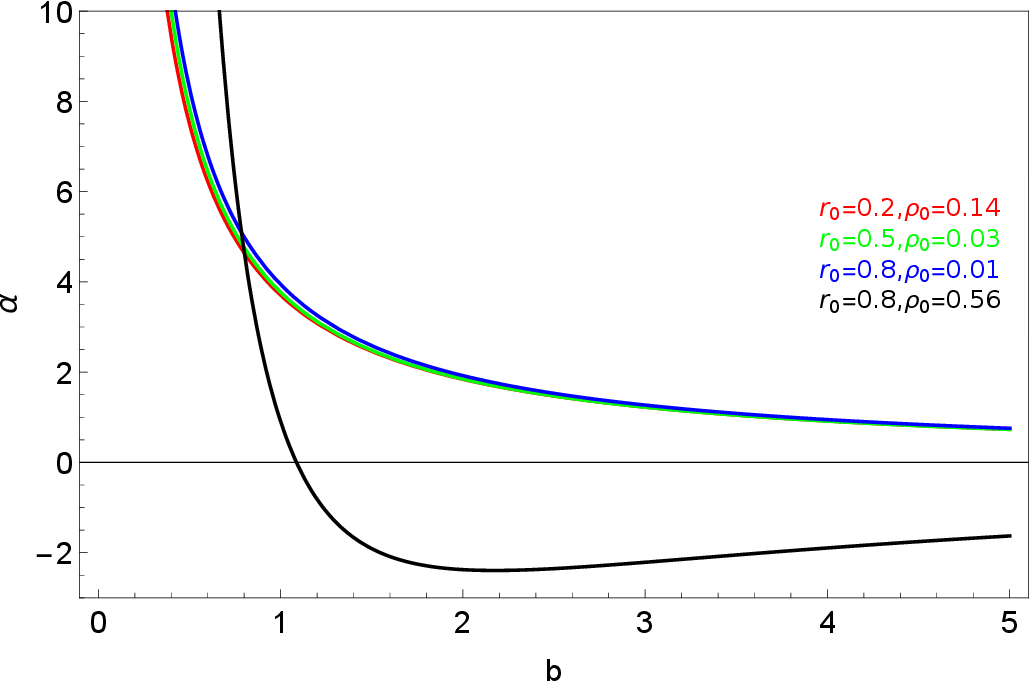}
    \caption{Profile of deflection angle $\alpha$ as a function of impact parameter $b$ for various $r_0$ and $\rho_0$ with fixed $r_s=2$. Right: $r_0$ and $\rho_0$ are chosen to be consistent with bounds from $M87^\ast$ and $SrgA^\ast.$ } 
  \label{Lensing}
\end{figure}

Figure \ref{Lensing} illustrates the deflection angle for several $r_0,\rho_0$ values. This plot shows the gravitational deflection angle as a function of the impact parameter for black holes surrounded by Dehnen $\left(1,4,\frac{3}{2}\right)$ dark matter profile with black hole mass $M = 1$. The results show that for low impact parameters $b \to 0$, the deflection angle increases significantly, indicating a strong gravitational lensing effect as one moves closer to the black hole. As the impact parameter grows, the deflection angle falls and approaches zero, which is expected. Based on the inequalities \eqref{negalpha}, we choose the parameters such that \eqref{negalpha} are satisfied to demonstrate negative deflection angle as a blue curve. In the right panel, we display deflection angle against impact parameter where the value of $r_0$ and $\rho_0$ agree with the bounds of Shadow radius discussed in the aforementioned subsection. Interestingly, even when the constraints from $M87^\ast$ and $SrgA^\ast$ are respected, the repulsive deflection angle remains visible. 

\section{Lyapunov Exponent}\label{sec:Lyapunov}
In this section, we investigate the Lyapunov exponent associated with the unstable circular orbits of test particles around the black hole embedded in the Dehnen dark matter halo. The Lyapunov exponent quantifies the rate of exponential divergence between neighboring trajectories and thus provides a direct measure of the orbital instability in the spacetime. Since the presence of the halo alters the geometry and the shape of the effective potential, the stability properties of circular orbits are expected to differ from those in the pure Schwarzschild case.

To analyze the instability of null geodesics, let us first recall the radial null geodesics from \eqref{radialnull}
\begin{align}
    \dot{r}^2 &= V_r(r)= E^2 -\frac{h(r)L^2}{r^2}. \label{Vrpotential}
\end{align}
Now let us consider the Euler-Lagrange equation
\begin{align}
    \frac{dp_q}{d\lambda} &= \frac{d}{d\lambda}\left(\frac{\partial\mathcal{L}}{\partial \dot{q}}\right) = \frac{\partial\mathcal{L}}{\partial q},
\end{align}
where $p_q$ is generalized momenta. By linearizing the equation of motion above around circular orbit of constant $r$, one may define an infinitesimal evolutionary matrix \cite{Mondal:2020uwp}
\begin{align}
    K_{ij} &= \begin{pmatrix}
0 & K_1 \\
K_2 & 0 
\end{pmatrix},
\end{align}
where 
\begin{align}
    K_1 &= \frac{d}{dr}\left(\dot{t}^{-1}\frac{\partial\mathcal{L}}{\partial r}\right), \\
    K_2 &= \frac{1}{\dot{t}g_{rr}}.
\end{align}
Principal Lyapunov exponent can be defined as \cite{Cardoso:2008bp}
\begin{align}
    \sigma^2 &= K_1 K_2,
\end{align}
For a circular geodesics, $V_r(r_c)=V'_r(r_c)=0$ \cite{Bardeen1972}. Thus, $K_1$ can be shown explicitly as
\begin{align}
    K_1 &= \left(\frac{1}{2g_{rr}\dot{t}}\left(g_{rr}^2V_r\right)'\right)'\bigg\rvert_{r=r_c}.
\end{align}
Here, we denote $'$ as derivative with respect to radial coordinate $r$. Thus, generic Lyapunov exponent for static spherically symmetric spacetime is obtained \cite{Mondal:2020uwp,Cardoso:2008bp}
\begin{align}
    \sigma &= \sqrt{\frac{V_r''}{2\dot{t}}}\Bigg\rvert_{r=r_c}.
\end{align}
Here, we consider only the positive sign of the Lyapunov exponent. From \eqref{Vrpotential}, the two conditions of circular null geodesics imply
\begin{align}
    \frac{E}{L} &= \pm \sqrt{\frac{h_c}{r_c^2}},~~~~\text{and}~~~~\frac{h'_c}{h_c} = \frac{2}{r_c}, \label{conds}
\end{align}
where $h(r_c)=r_c$. Finally, the Lyapunov exponent can be expressed explicitly as 
\begin{align}
    \sigma &= \sqrt{h_c\left(\frac{h_c}{r_c^2}-\frac{h''_c}{2}\right)}, \label{lyapunov0} \\
    &= \sqrt{\frac{\left(e^{\Delta}r_c-r_s\right)\left[\left(2r_c^2-4M_{DM}^2\right)\left(r_0+r_c\right) +M_{DM}r_c\left(r_0+4r_c\right)\right]}{2e^{-\Delta}\left(r_0+r_c\right)r_c^5}}. \label{lyapunov}
\end{align}
For the sake of presentation, we simplify $\sigma$ by using $M_{DM}$ rather than the usual $\rho_0$ and introduce $\Delta\equiv \frac{4M_{DM}(r_0+r_c)}{r_0r_c}$. In the absence of dark matter, $e^{\pm\Delta}=1$, the Lyapunov exponent \eqref{lyapunov} reduce to $\sigma = \sqrt{\frac{r_c-r_s}{r_c^3}}=\frac{1}{3\sqrt{3}}$ which agrees with those of the Schwarzschild case \cite{Mondal:2020uwp}. The Lyapunov exponent denotes the inverse of the instability timescale between neighboring geodesics. The circular null geodesics is unstable when $\sigma>0$. Clearly, \eqref{lyapunov} being real number signifies the instability of geodesics. 

\begin{figure}[H]
    \centering
    \includegraphics[scale=0.35]{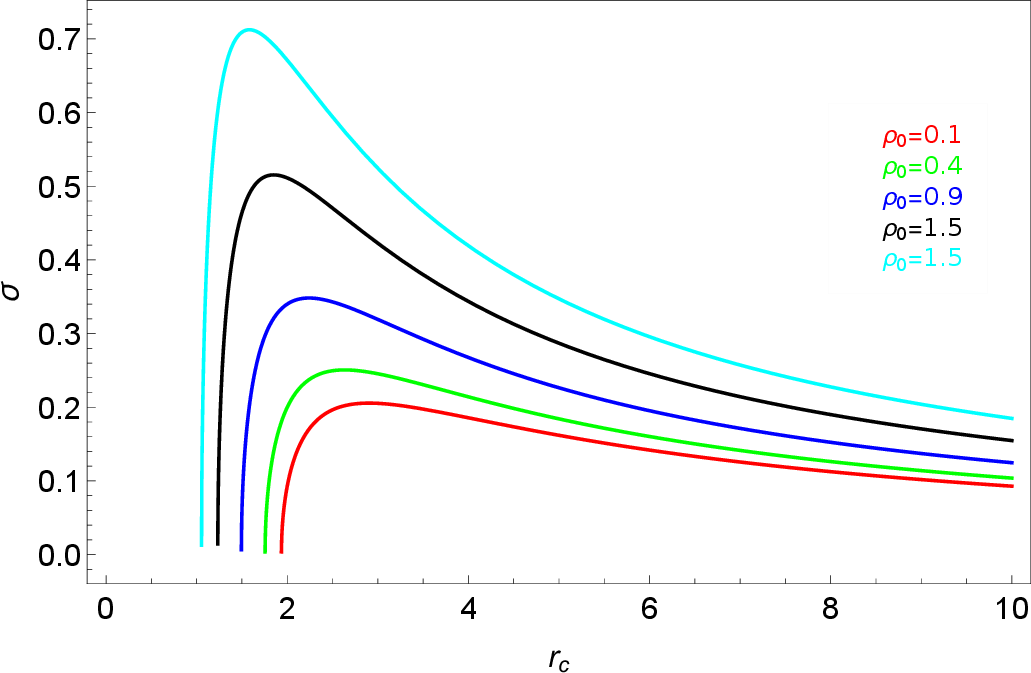}
    \includegraphics[scale=0.35]{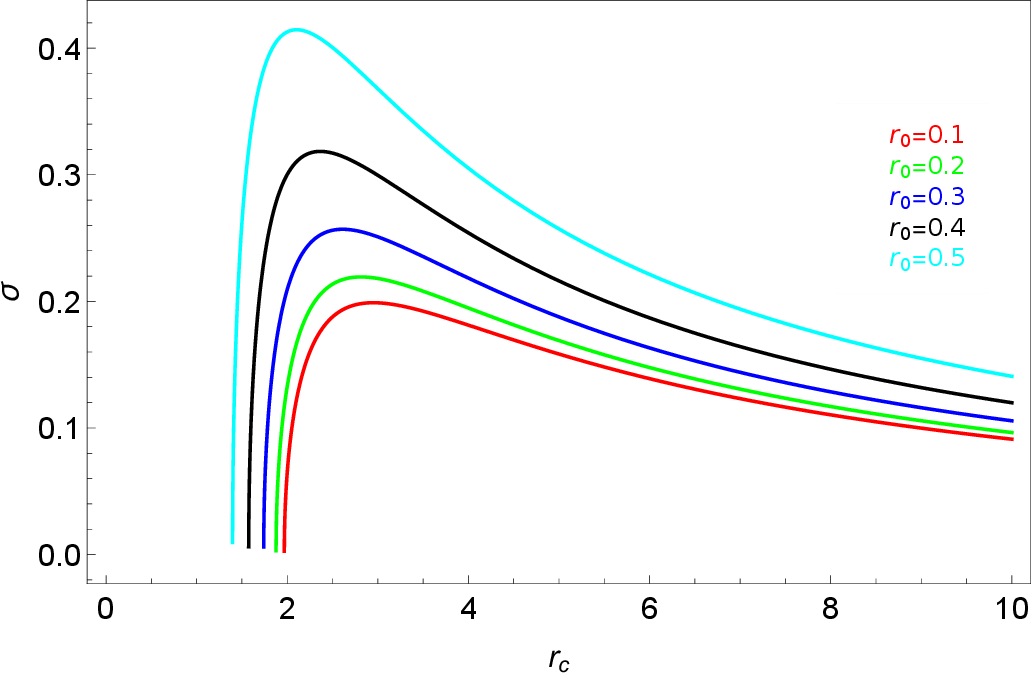}
    \caption{The Lyapunov exponent as a function of $r_c$ for various value of $\rho_0$ and $r_0$. Left: fixed $r_0=0.1$ Right: fixed $\rho_0=0.05$.} 
  \label{fig:lyapunov}
\end{figure}

As a demonstration, we display behaviour of the Lyapunov exponent $\sigma$ in Fig.~\ref{fig:lyapunov} for several cases of $\rho_0$ and $r_0$.
For each case, there exists a minimum value of $r_c$ for which the Lyapunov exponent remains as real number. For instance, the Lyapunov exponent becomes complex number in $0\leq r_c \leq 1.93$ for $\rho_0=r_0=0.1$. We observe that increases in $\rho_0$ and $r_0$ generally lead to an overall increase in the magnitude of $\sigma$.

\section{Eikonal Quasinormal Modes}\label{sec:QNMs}
Quasinormal modes (QNMs) characterize the damped oscillations of perturbations around a black hole or compact object, representing the system’s natural response to external disturbances. Their complex frequencies have real parts corresponding to oscillation frequencies and imaginary parts describing decay rates, providing valuable insight into the stability and dynamical properties of the spacetime. 

Let us consider a massive relativistic {massless}  bosonic field perturbing the spherically symmetric black hole solution embedded in a Dehnen $\left(1, 4, \frac{3}{2}\right)$ dark matter halo. The dynamics of the  field are governed by the covariant Klein-Gordon equation, which in curved spacetime takes the form
\begin{gather}
\frac{1}{\sqrt{-g}} \partial_\mu \left(\sqrt{-g} g^{\mu \nu} \partial_\nu \right)  \psi = 0,
\end{gather}
where $m$ is the mass of the bosonic field and $g$ is the determinant of the metric tensor. 

By substituting the explicit form of the metric given in \eqref{metric} and following the procedure outlined in \cite{Senjaya:2024rse}, we exploit the spherical symmetry of the spacetime and adopt the separation ansatz
\begin{gather}
\psi(t,r,\theta,\phi) = e^{-i\omega t} \frac{R(r)}{r} Y_\ell^{m_\ell}(\theta,\phi),
\end{gather}
where $Y_\ell^{m_\ell}(\theta,\phi)$ are the standard spherical harmonics,

Using this ansatz, the Klein-Gordon equation reduces to a radial wave equation for $R(r)$,
\begin{align}
\frac{d^2R}{dr_\ast^2}+\left[\omega^2-\frac{h}{r^2}\left\{\ell(\ell+1)+rh'\right\}\right] R=0, \label{radial}
\end{align}
where we have defined the tortoise coordinate
\begin{equation}
dr_\ast = \frac{dr}{h(r)}.
\end{equation}
Equation \eqref{radial} describes the radial propagation of scalar perturbations in the black hole–Dehnen halo spacetime, with the metric function $h(r)$ encoding the influence of both the central black hole and the surrounding dark matter distribution. Solving this equation with appropriate boundary conditions, i.e., purely ingoing wave at the horizon and purely outgoing wave at asymptotically infinity, allows us to determine the quasinormal mode frequencies. As a result, the corresponding frequencies are in the form of complex frequencies, i.e., $\omega=\omega_R\pm i\omega_I$.

To obtain the frequencies, there are several numerical approaches. However, there is an approximation that provides a useful analytical formula of quasinormal modes. We can consider geometric-optics or eikonal limit as proposed in \cite{Ferrari:1984zz,Mashhoon:1985cya,Mashhoon:1985b}. In the eikonal limit, the QNM spectrum admits a clear geometric interpretation: the real part of the frequency corresponds to the angular velocity of the unstable circular photon orbit, while the imaginary part is determined by the Lyapunov exponent associated with the orbit’s instability \cite{Cardoso:2008bp}. This correspondence establishes a direct link between the classical motion of photons and the linear perturbative dynamics of the black hole. Analyzing eikonal QNMs therefore allows us to investigate how the parameters of the Dehnen halo influence both the oscillatory behavior and damping of perturbations in the black hole–dark matter system. 

In the eikonal limit, where $l \gg 1$, the radial equation \eqref{radial} simplifies considerably. Neglecting terms of order $\mathcal{O}(1)$ compared to $l^2$, we obtain,
\begin{gather}
  {\partial }_{r_\ast}^2  \mathcal{R}(r_\ast)+W(r)\mathcal{R}(r_\ast)=0,\\
  W(r)=\omega^2-\frac{h}{r^2}l^2=\omega^2-l^2V_{eff}.
\end{gather}
To determine the quasinormal mode (QNM) frequencies in the eikonal regime, we employ the analytical first order WKB approximation. This semi-analytical method provides an efficient and reliable way to estimate the complex frequencies associated with the effective potential barrier. The standard Iyer–Will WKB quantization condition is given by \cite{Schutz:1985km},
\begin{equation}
    \frac{W(r_0)}{\sqrt{2W^{(2)}(r_0)}}=-i\left(n+\frac{1}{2}\right), \label{WKB}
\end{equation}
where $n=0,1,2,\dots$ denotes the overtone number and,
\begin{equation}
   W^{(2)}(r_0)=\left(\frac{d^2 W}{d{r^2_\ast}}\right)_{r=r_0}.
\end{equation}
The point $r_0$ corresponds to the extremum of both $W(r)$ and the effective potential $V_{eff}(r)$, consequently, $r_0=r_{ph}$ is physically interpreted as the radius of the circular null orbit (or the photon sphere). 
The explicit expression of the WKB formula in \eqref{WKB} can be rewritten explicitly as
\begin{align}
    \omega^2_{QNM} &= \ell^2 V_{eff}(r_{ph}) - i\left(n+\frac{1}{2}\right)\sqrt{-2\ell^2 V_{eff}^{(2)}(r_{ph})}. 
\end{align}
After taking square root and expanding for large $\ell$, we obtain quasinormal frequency in a more transparent form as,
\begin{equation}
    \omega_{QNM}\approx \ell\sqrt{V_{eff}(r_{ph})}-i\left(n+\frac{1}{2}\right)\sqrt{-\frac{V_{eff}^{(2)}(r_{ph})}{2V_{eff}(r_{ph})} }. \label{WKBeikonal}
\end{equation}
Now, let us evaluate the second derivative of the effective potential with respect to the tortoise coordinate. This is given by
\begin{align}
\left(\frac{d^2 V_{eff}}{dr^2_\ast}\right)_{r=r_{ph}}= \left[h(r)\frac{d}{dr}\left(h(r)\frac{dV_{eff}}{dr}\right)\right]_{r=r_{ph}}
&= h^2(r_{ph}) V''_{eff}(r_{ph})
\end{align}
where $dV_{eff}/dr_\ast=0\rvert_{r=r_{ph}}$ is implemented. We find that $V_{eff}$ is maximum at a radius satisfying $h'_{ph}=2h_{ph}/r_{ph}$. By comparing with \eqref{conds}, we observe that  $r_{ph}$ coincides with circular null orbit $r_c$. Then, it is useful to consider
\begin{align}
    -\frac{V_{eff}^{(2)}(r_c)}{2V_{eff}(r_c)} &= -\frac{1}{2}h_c\left(h''_c - \frac{2h_c}{r_c^2}\right).
\end{align}
In comparison with \eqref{WKBeikonal}, this allows us to rewrite \eqref{WKBeikonal} in term of the Lyapunov exponent
\begin{align}
\omega_{QNM} &= \ell\sqrt{\frac{h_c}{r_c^2}}-i\left(n+\frac{1}{2}\right)\sigma. \label{eikformula}
\end{align}
Interestingly, the real part of eikonal formula above has a connection to photon's critical impact parameter $b_c$ and then to black hole's shadow radius via \eqref{bcrit} and \eqref{Rs}. It is also important to stress out that the imaginary part of the black hole’s quasinormal mode spectrum is directly proportional to $\sigma$, which quantifies the stability or instability of the corresponding classical photon circular orbit. For a stable circular orbit, where $\sigma^2 < 0$, the quasinormal mode frequency $\omega_{\rm QNM}$ is entirely real, indicating undamped oscillations. In contrast, an unstable circular orbit, for which $\sigma^2 > 0$, yields a complex $\omega_{QNM}$ with a negative imaginary part, corresponding to decaying perturbations and reflecting the intrinsic instability of the orbit.

For special case with $\rho_0 r_0^3\ll 1$, we can perform series expansion and obtain,
\begin{equation}
 \omega_{QNM}= \frac{2\ell}{3\sqrt{3}\, r_s^2}\Bigl[ r_s -\frac{16}{3}\pi\rho_0 r_0^2 (r_0 - 3 r_s)\Bigr]-i\frac{2\left(n+\frac{1}{2}\right)}{3\sqrt{3}\, r_s^2}\Bigl[r_s - \frac{16}{3}\pi\rho_0 r_0^2 (r_0 - 4 r_s)\Bigr]
\end{equation}
For the Schwarzschild case, where $V_{eff} = \frac{1}{r^2} \left( 1 - \frac{2M}{r} \right)$ and $r_{ph} = 3M$, it is straightforward to compute $\sigma = \frac{1}{9M^2}$ and $V_{eff}(r_{ph}) = \frac{1}{27M^2}$. Substituting these values into the eikonal QNM formula, one recovers the well-known expression \cite{Ferrari:1984zz},
\begin{equation}
    \omega_{QNM}= \frac{\ell}{3\sqrt{3}M}-\frac{i\left(n+\frac{1}{2}\right)}{3\sqrt{3}M}.
\end{equation}

This result clearly demonstrates how the general eikonal QNM expression reduces to the standard Schwarzschild limit, providing a useful consistency check for calculations involving more general black hole spacetimes, such as those embedded in a Dehnen dark matter halo.

\section{Conclusion}\label{sec: conclu} 
In this work, we construct and analyze a new static, spherically symmetric black hole solution embedded in a Dehnen $\left(1, 4, \frac{3}{2}\right)$ dark matter halo. The Dehnen profile has been widely used in astrophysics to model elliptical galaxies and bulges, but its influence on black hole solutions has yet to be fully investigated. By embedding a Schwarzschild black hole in this density distribution, we can obtain an exact analytic form of the metric and investigate how the halo affects spacetime structure. The halo parameters, $\rho_0$ and $r_0$, have a significant impact on the horizon location and curvature invariants, while maintaining the essential singularity at the origin, according to the mass distribution and metric functions.

We then look at the energy conditions of the combined black hole-halo spacetime. Our findings indicate that, while the strong energy condition is generally met, both the weak and dominant energy conditions are violated. Some combinations of $\rho_0$ and $r_0$ can satisfy the null energy condition, but only outside the horizon. These findings highlight the nontrivial behavior of the energy-momentum tensor produced by the Dehnen halo, implying that the halo's effective stress-energy may mimic exotic matter properties in curved spacetime.

The thermodynamic properties of the system are also investigated. We construct the black hole mass function and derive analytical functions for entropy, Hawking temperature, heat capacity and Gibbs free energy. Unlike the Schwarzschild case, which is always thermodynamically unstable, the presence of a Dehnen halo creates regions with positive heat capacity, indicating local stability. Furthermore, the system experiences second-order phase transitions, as evidenced by discontinuities in heat capacity while entropy remains constant. The Gibbs free energy analysis also reveals that higher halo density and larger core radius improve the global stability of the system by minimizing the positive Gibbs free energy region, demonstrating that dark matter can act as a stabilizing medium for black holes.

We explore the observable shadow of the black hole encircled by the Dehnen halo and null geodesics. Depending on the halo parameters, the photons' effective potential shows changes to the photon sphere and circular orbits. The apparent shadow radius and the critical impact parameter are directly impacted by these modifications. Our computations indicate that the shadow size grows with scale radius and halo density, pointing to possible dark matter observational signatures in black hole imaging. Additionally, weak gravitational lensing is examined, demonstrating that the halo considerably modifies light deflection at low impact parameters. These findings provide a useful framework for relating theoretical models with astrophysical observations by demonstrating that embedding a black hole in a Dehnen-type dark matter halo results in unique thermodynamic and optical signatures.

Furthermore, we clarify how the dynamics of photon orbits are reflected in the spacetime’s response to perturbations. We analytically show that the Lyapunov exponent describing the instability of circular photon orbits is directly linked to the imaginary part of the massless quasinormal mode frequencies in the eikonal limit, which set the decay times of relativistic perturbations. We find that increasing both $\rho_0$ and $r_0$ increases the divergence rate of two nearby orbits. This result emphasizes the photon sphere as the key structure connecting black hole shadows, gravitational lensing and quasinormal ringing. Consequently, when a Schwarzschild black hole is embedded in the Dehnen $\left(1, 4, \frac{3}{2}\right)$ dark matter halo, changes in the photon orbit instability leave simultaneous imprints on both optical observables and the damping of perturbations, offering a complementary way to probe the surrounding dark matter environment.

As a further extension, a comprehensive analysis of the optical appearance of a black hole embedded in a Dehnen dark matter halo would be highly valuable. In particular, calculating the resulting shadow, lensing features, and associated observables in this composite spacetime could provide an astrophysical probe of the Dehnen halo parameters. Such a study may help constrain the structure and distribution of Dehnen-type dark matter halos through comparison with high-resolution imaging and lensing data. Research in this direction is currently underway.


\section*{Acknowledgment}
TC and SP acknowledge funding support from the NSRF via the Program Management Unit for Human Resource and Institutional Development, Research and Innovation grant number $B39G680009$.


\bibliography{sn-bibliography}


\end{document}